\documentclass[fleqn,usenatbib]{mnras}

\usepackage{newtxtext,newtxmath}

\usepackage[T1]{fontenc}

\DeclareRobustCommand{\VAN}[3]{#2}
\let\VANthebibliography\thebibliography
\def\thebibliography{\DeclareRobustCommand{\VAN}[3]{##3}\VANthebibliography}
\usepackage{graphicx}	
\usepackage{amsmath}	
\usepackage{xfrac,nicefrac}
\usepackage{subcaption}

\newcommand{\Mdot}{\mbox{\,$\rm M_{\odot}$}}        
\newcommand{\Zdot}{\mbox{\,$\rm Z_{\odot}$}}        
\newcommand{\aov}{$\alpha_{\rm ov}$}		
\newcommand{\fov}{$f_{\rm ov}$}		
\newcommand{\amlt}{$\alpha_{\rm mlt}$}		

\title[VMS wind yields]{Stellar Wind Yields of Very Massive Stars}
\author[E. R. Higgins et al.]{
Erin R. Higgins$^{1}$\thanks{E-mail: erin.higgins@armagh.ac.uk},
Jorick S. Vink$^{1}$,
Raphael Hirschi$^{2,3}$,
Alison M. Laird$^{4}$,
Gautham N. Sabhahit$^{1}$\\
$^{1}$Armagh Observatory and Planetarium, College Hill, Armagh BT61 9DG, N. Ireland\\
$^{2}$Astrophysics Group, Keele University, Keele, Staffordshire ST5 5BG, UK\\
$^{3}$Kavli Institute for the Physics and Mathematics of the Universe (WPI),\\ University of Tokyo, 5-1-5 Kashiwanoha, Kashiwa 277-8583, Japan\\
$^{4}$School of Physics, Engineering and Technology, University of York, York, YO10 5DD, UK}

\date{Accepted. Received}
\pubyear{2023}

\begin{document}

\maketitle
\begin{abstract}
\noindent
The most massive stars provide an essential source of recycled material for young clusters and galaxies. While very massive stars (VMS, M$>$100\Mdot) are relatively rare compared to O stars, they lose disproportionately large amounts of mass already from the onset of core H-burning. VMS have optically thick winds with elevated mass-loss rates in comparison to optically thin standard O-star winds. 
We compute wind yields and ejected masses on the main sequence, and we compare enhanced mass-loss rates to standard ones. We calculate solar metallicity wind yields from MESA stellar evolution models in the range 50 -- 500\Mdot, including a large nuclear network of 92 isotopes, investigating not only the CNO-cycle, but also the Ne-Na and Mg-Al cycles. VMS with enhanced winds eject 5-10 times more H-processed elements (N, Ne, Na, Al) on the main sequence in comparison to standard winds, with possible consequences for observed anti-correlations, such as C-N and Na-O, in globular clusters. We find that for VMS 95\% of the total wind yields is produced on the main sequence, while only $\sim$ 5\% is supplied by the post-main sequence. This implies that VMS with enhanced winds are the primary source of $^{26}$Al, contrasting previous works where classical Wolf-Rayet winds had been suggested to be responsible for Galactic $^{26}$Al enrichment. Finally, 200\Mdot\ stars eject 100 times more of each heavy element in their winds than 50\Mdot\ stars, and even when weighted by an IMF their wind contribution is still an order of magnitude higher than that of 50\Mdot\ stars.

\end{abstract}


\section{Introduction}\label{intro}

The chemical composition of galaxies relies on the production of elements in stars, which are subsequently released in stellar winds and supernovae. This ejected material is then responsible for enriching the neighbouring environment with heavier elements. The evolution of galaxies therefore depends on the main production sites of various chemical isotopes and the relevant feedback of enriched material which ultimately affects the metallicity of a given galaxy or cluster \citep{tinsley80}. The origin of elements thereby concerns the stellar nucleosynthesis, wind ejecta and chemical yields, of a given population, providing a broad perspective on galactic chemical evolution \citep[GCE;][]{Kob20}. 

This rejuvenation of galaxies over generations of stars has led us to the metal-rich environment of our own Galaxy, with abundant quantities of carbon (C), oxygen (O), nitrogen (N) and iron (Fe). These fusion products are key for establishing life in our modern-day Universe, including the enrichment of our solar system with elements such as radioactive aluminium ($^{26}$Al). Surveys of the Milky Way by COMPTEL \citep{diehl95} and INTEGRAL \citep{diehl06} found $\sim$ 3\Mdot\ of $^{26}$Al which is a radioactive isotope of Al with a half-life of $\sim$ 0.7Myr. The observed $^{26}$Al was therefore produced and expelled into our Galaxy recently, likely by massive stars.

Another intriguing puzzle from the last few decades concerns the origin of the C-N and Na-O anti-correlations in Globular Clusters \citep{bastianlardo18}. 
These hydrogen-burning by-products have been suggested to either originate from asymptotic giant branch (AGB) stars \citep{Ercole10}, massive stars \citep{Decressin}, and even supermassive stars (SMS) \citep{Denissenkov14,Gieles18}. However, it remains to be shown that SMS are actually formed in Nature.
Very massive stars (VMS) with masses over 100$\Mdot$ have therefore been proposed as an alternative polluter \citep{vink18} as VMS are actually seen in Nature, such as in the Arches cluster of the Milky Way and the Tarantula Nebula of the Large Magellanic Cloud (LMC). 

These VMS have been theoretically predicted, and observed, to have enhanced stellar winds, expelling significant amounts of mass during their lifetime. \cite{vink11} calculated Monte Carlo simulations of VMS finding an upturn in the mass-loss rates of stars above a given transition point where stellar winds change from being optically thin to optically thick. A similar increase in the mass-loss rates was observed for the VMS or Hydrogen-rich Wolf-Rayet class of WNh stars in the LMC by \cite{best14}. Recent work by \cite{Sabh22} provided a physically-motivated wind prescription for VMS which adopts enhanced winds above the \cite{VG12} transition point, whilst retaining standard O-star rates for stars below the transition point. 
A comparison of the observed VMS in both the Galaxy and LMC showed good agreement with stellar properties such as luminosities (L) and effective temperatures (T$_{\rm{eff}}$). A subsequent study by \cite{higgins22} implemented this new wind prescription for stars with initial masses ranging from 100\Mdot\ to 1000\Mdot\ discovering that the surface Hydrogen (H) abundance could be used to infer the interior H-burning abundance. This is a result of chemically-homogeneous evolution (CHE), where even non-rotating stars are fully mixed as a result of the large convective cores of VMS which comprise $\sim$ 90\% of the entire star. 

With such enhanced winds already at the Zero-Age-Main-Sequence (ZAMS), VMS could be the main contributors of processed material which regenerates their host young cluster. In fact, while supernovae ejecta likely dominate the contribution of massive star yields for M $\sim$ 20\Mdot, this would only occur after $\sim$ 5-10Myr, while the constant source of enriched material from VMS likely dominates the first 1-5Myr of a given region. For this reason, we explore the contribution of VMS wind yields at solar metallicity (Z) for a range of masses, implementing the enhanced wind prescription for VMS. We compare the effects of adopting the optically thin O star winds which have been previously implemented in the literature for VMS. Ejected masses and net wind yields are provided for VMS on the main sequence (MS) and we explore the effects of MS winds on the post-MS. Finally, we examine the contribution of VMS winds when weighted by an initial mass function (IMF) as compared to standard O stars. The wind and supernovae yields of $^{26}$Al from massive single stars have been estimated previously by \cite{limongi06,limongi18, martinet22}, while the effect of binary interaction has been explored more recently by \cite{brinkman19,brinkman21}.  

We present an overview of our model grid at solar metallicity in Sect. \ref{method}, with a description of stellar winds in Sect. \ref{sectmassloss}. The results of our stellar models are shown in Sect. \ref{sectresults}, with details of VMS nucleosynthesis in Sect. \ref{sectnucleo}, key features of VMS evolution in Sect. \ref{sectevolution} and observable surface enrichment in Sect.\ref{observation}. We provide the ejected masses and yield calculations for various isotopes in Sect. \ref{sectyield}, with the contribution of MS mass loss discussed in Sect. \ref{sectMS}, and a discussion of the impact that MS winds have on the post-MS following in Sect. \ref{sectpostMS}. 
A comparison of VMS and O star wind yields is provided in Sect. \ref{lowmass} as a function of their contribution to their host environment for a given IMF. Finally, we provide our conclusions in Sect. \ref{conclusions}.

\section{Method}\label{method}
In this section, we provide an overview of our evolutionary models with the relevant wind prescriptions and nuclear reactions necessary for estimating wind yields at solar Z. We compare two stellar wind prescriptions in order to showcase the impact of VMS wind yields on their environment. In Sect. \ref{sectyield} we describe our method of calculating ejected masses and net yields for a given initial mass and chemical isotope.
\begin{table}
    \centering
    \begin{tabular}{c|c|c|c}
    \hline
      Isotope &  Mass fraction & Isotope & Mass fraction\\
      \hline \hline
      $^{1}$H   & 0.719986 & $^{20}$Ne & 1.356E-3\\
       $^{2}$H & 1.440E-5 & $^{22}$Ne & 1.097E-4 \\
        $^{3}$He  & 4.416E-5 & $^{23}$Na & 2.9095E-5\\
         $^{4}$He & 0.266 & $^{24}$Mg & 4.363E-4\\
          $^{12}$C & 2.380E-3 &$^{25}$Mg & 5.756E-5\\
          $^{14}$N & 7.029 E-4 & $^{26}$Mg & 6.585E-5\\
        $^{16}$O & 6.535E-3 & $^{27}$Al & 5.051E-5 \\
        $^{18}$O & 1.475E-5 & $^{28}$Si & 5.675E-4\\
        $^{19}$F & 3.475E-7 & $^{32}$S & 2.917E-4\\
        \hline
    \end{tabular}
    \caption{Initial abundances of chemical isotopes in mass fractions for our grid of models at \Zdot.}
    \label{tab:abundances}
\end{table}
\begin{figure*}
\begin{subfigure}{.5\textwidth}
  \centering
  \includegraphics[width=.95\linewidth]{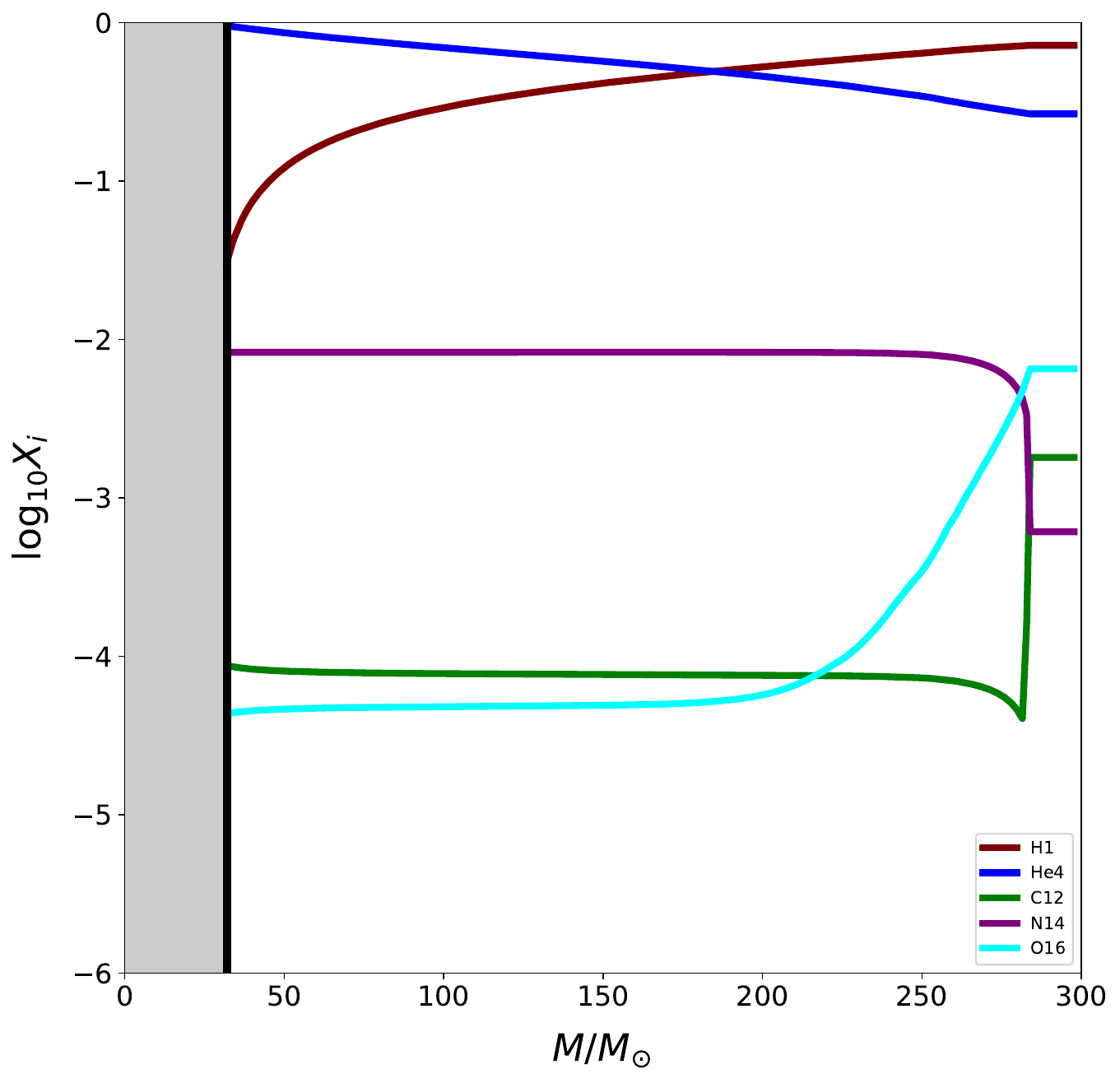}
  \caption{Enhanced V11 optically thick winds for VMS}
  \label{fig:sfig1}
\end{subfigure}%
\begin{subfigure}{.5\textwidth}
  \centering
  \includegraphics[width=.95\linewidth]{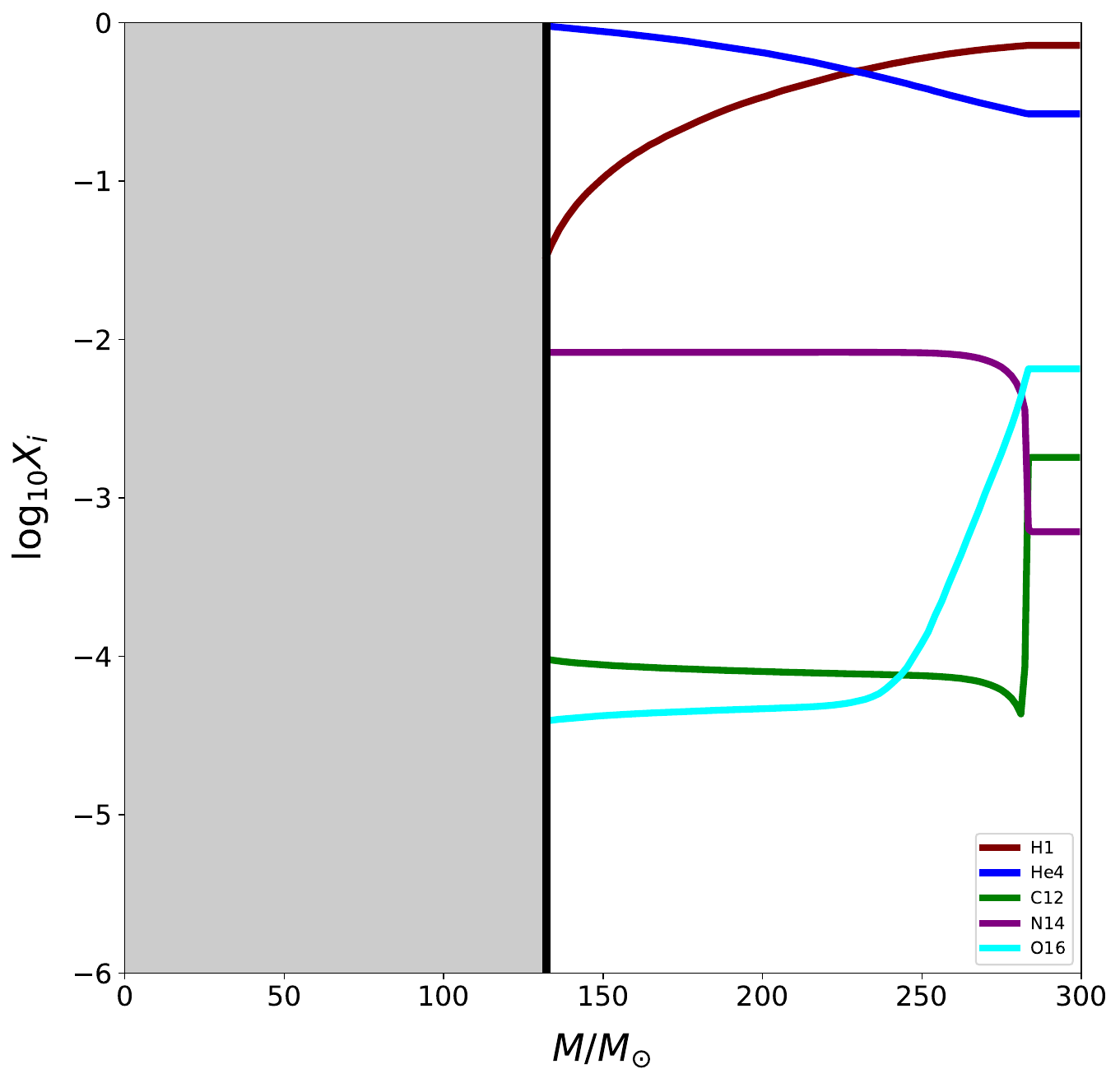}
  \caption{Standard V01 optically thin O star winds applied to VMS}
  \label{fig:sfig2}
\end{subfigure}
\caption{Time evolution of the surface abundance of H, He, C, N and O as a function of mass, for 300\Mdot\ models at \Zdot. As the star loses mass the time evolution goes from right to left. We apply the optically-thick, enhanced wind prescription outlined in equation \eqref{V11} on the left, and the standard optically thin, O star wind prescription from equation \eqref{V01} on the right. The white region shows the mass lost over the MS lifetime, while the grey shaded region illustrates the remaining terminal age main-sequence (TAMS) stellar mass with the black line denoting the TAMS. }
\label{fig:V01V11compare}
\end{figure*}
\subsection{Stellar models}\label{sectmodels}
Stellar models have been calculated using the one-dimensional stellar evolution code \texttt{MESA} \citep[r10398;][]{Pax11,Pax13,Pax15,Pax18,Pax19} for a grid of initial masses of 100, 200, 300, 400, and 500\Mdot. We have also computed comparable models at lower initial masses of 50\Mdot\ and 80\Mdot. All calculations begin with a pre-main sequence and then evolve from the ZAMS until core O-exhaustion ($^{16}\rm{O}_{\rm{c}}$ $<$ 0.00001). We implement a nuclear reaction network which includes the relevant isotopes for massive star evolution until the end of core O-burning. This nuclear network comprises the following 92 isotopes: n, $^{1, 2}$H, $^{3, 4}$He, $^{6, 7}$Li, $^{7, 9, 10}$Be, $^{8, 10, 11}$B, $^{12, 13}$C, $^{13-16}$N, $^{14-19}$O, $^{17-20}$F, $^{18-23}$Ne, $^{21-24}$Na, $^{23-27}$Mg, $^{25-28}$Al, $^{27-33}$Si, $^{30-34}$P, $^{31-37}$S, $^{35-38}$Cl, $^{35-41}$Ar, $^{39-44}$K, and $^{39-44,46, 48}$Ca. Our stellar models are computed with solar metallicity, where $X$ $=$ 0.720, $Y$ $=$ 0.266 and \Zdot\ $=$ 0.014 where the relative composition is adopted from \cite{asplund09}, provided in Table \ref{tab:abundances}. We avail of the OPAL opacity tables from \cite{RogersNayfonov02}, and adopt nuclear reaction rates from the JINA Reaclib Database \citep{Cyburt10}. 

The mixing-length-theory (MLT) of convection describes the treatment of convection in our models, where we apply an efficiency of \amlt $=$ 1.67 \citep{arnett19}. The Schwarzschild criterion defines the convective boundaries in our models and as such we do not implement semiconvective mixing. For convective boundary mixing (CBM), we include 
the exponential decaying diffusive model of \citet{Freytag1996} \citep[see also][]{Herwig2000}
with \fov $=$ 0.03 (corresponding to \aov $\simeq$ 0.3) for the top of convective cores and shells, and with \fov $=$ 0.006 for the bottom of convective shells.
\cite{bowman20review} find a range of \aov\ from asteroseismology results with \aov\ up to 0.4, so our value for the top of convective zones falls in the range of asteroseismology-inferred values. This value also falls in between the majority of published large grids of stellar models such as $\alpha_{\textrm{ov}}=0.1$ in \citet[][]{Ekstroem2012}, $\alpha_{\textrm{ov}}$ $=$ 0.335 in \citet[][]{Brott2011}, and recent studies on CBM \citep[][]{Higgins, Scott21} supporting values for \aov\ up to at least 0.5 for stars above 20\,M$_\odot$.
For the bottom boundary, a CBM value of 1/5 the value of the top boundary is based on 3D hydrodynamic simulations \citep{Cristini2017,Cristini2019,Rizzuti2022} finding that CBM is slower at the bottom boundaries due to being stiffer and therefore harder to penetrate.

In order to evolve such high mass models, without enhanced winds for comparisons, we apply convection in superadiabatic layers via the \texttt{MLT++} prescription which aids convergence of such models to late evolutionary stages. The temporal resolution of our models has been set with \texttt{varcontrol}\texttt{target} $=$ 0.0001, and a corresponding spatial resolution of \texttt{mesh}\texttt{delta} $=$ 0.5.

\subsection{Mass Loss}\label{sectmassloss}
In this work, we compare 2 stellar wind prescriptions, and explore their effects on VMS evolution and corresponding wind yields. Theoretical mass-loss rates of massive stars were calculated by \cite{Vink01} as a function of mass, luminosity, effective temperature, terminal velocity, and metallicity, 
\begin{equation}
\begin{split}
\mathrm{log} \;\dot{M}_\mathrm{V01} = \;\;& -6.697 \\ & + 2.194\;\mathrm{log}(L/L_\odot/10^5) \\ & -1.313\;\mathrm{log}(M/M_\odot/30)\\ & -1.226\;\mathrm{log}((\varv_\infty/\varv_{\mathrm{esc}})/2) \\ & +0.933\; \mathrm{log}(T_{\mathrm{eff}}/40000) \\ & -10.92\;\{\mathrm{log}(T_{\mathrm{eff}}/40000)\}^2 \\
&  + 0.85 \;\mathrm{log}(Z/Z_\odot)
\end{split}
\label{V01}
\end{equation}
as shown in equation \eqref{V01}. 
These mass-loss rates were calculated with Monte-Carlo (MC) simulations which trace the number of photons travelling below the photosphere through the stellar wind, thereby calculating the radiative acceleration and mass-loss rate. 
The MC simulations were calculated for hot (log$_{10}$ (T$_{\rm{eff}}$/K) $\geq$ 4.0), optically thin OB stars. This mass-loss recipe has been implemented across many stellar evolution and population synthesis codes for massive stars, and in some cases extrapolated to higher masses, which have been shown to under-predict the stellar winds of VMS, \citep{vink06,best14}.

Following this, \cite{vink11} computed MC simulations up to 300\Mdot\ finding a `kink' or upturn in the mass-loss rates at the highest masses. Similarly, the massive star observations in 30 Dor also displayed a `kink' in the mass-loss rates of the most massive stars \citep{best14}. This transition point aligns with the observed spectral transition from O stars with optically thin winds to Of/WNh stars with optically thick winds. While new dependencies on L/M were provided by \cite{vink11}, showing a strong $\Gamma$-dependence on $\dot{M}$, absolute rates were not calculated. As a result, a recent study by \cite{Sabh22} provided a complete mass-loss prescription which switches from the optically thin \cite{Vink01} rates below the transition point ($\sim$ 77\Mdot\ at \Zdot) to the updated \cite{vink11} rates for VMS above the transition point. Since there are a number of transition stars in the Arches cluster of the Milky Way and 30 Dor in the Large Magellanic Cloud (LMC), the absolute rates of both recipes were anchored to the transition point such that a complete recipe was achievable, 
\begin{equation}
\begin{split}
\mathrm{log}\; \dot{M}_\mathrm{V11} = \;\;& -8.445 \\ &+ 4.77\;\mathrm{log}(L/L_\odot/10^5)\\ & -3.99\;\mathrm{log}(M/M_\odot/30)\\ & -1.226\;\mathrm{log}((\varv_\infty/\varv_{\mathrm{esc}})/2) \\ 
&  + 0.761 \;\mathrm{log}(Z/Z_\odot)
\end{split}
\label{V11}
\end{equation}
as shown in equation \eqref{V11}. The \cite{Sabh22} study provided a comparison with observed transition stars and VMS in the Galaxy and LMC, showing excellent agreement with absolute mass-loss rates and evolutionary traits. In fact, the self-regulatory behaviour found in the enhanced wind VMS models, illustrated in the Hertzsprung-Russell diagram (HRD), demonstrated that observed VMS could be reproduced in a narrow effective temperature range with a steep drop in luminosity, as a result of the enhanced wind prescription. We adopt this mass-loss recipe throughout the paper, hereafter `V11'. We compare the updated V11 mass-loss recipe with the standard O star wind of \cite{Vink01}, hereafter `V01'.
\begin{figure}
    \centering
    \includegraphics[width = \columnwidth]{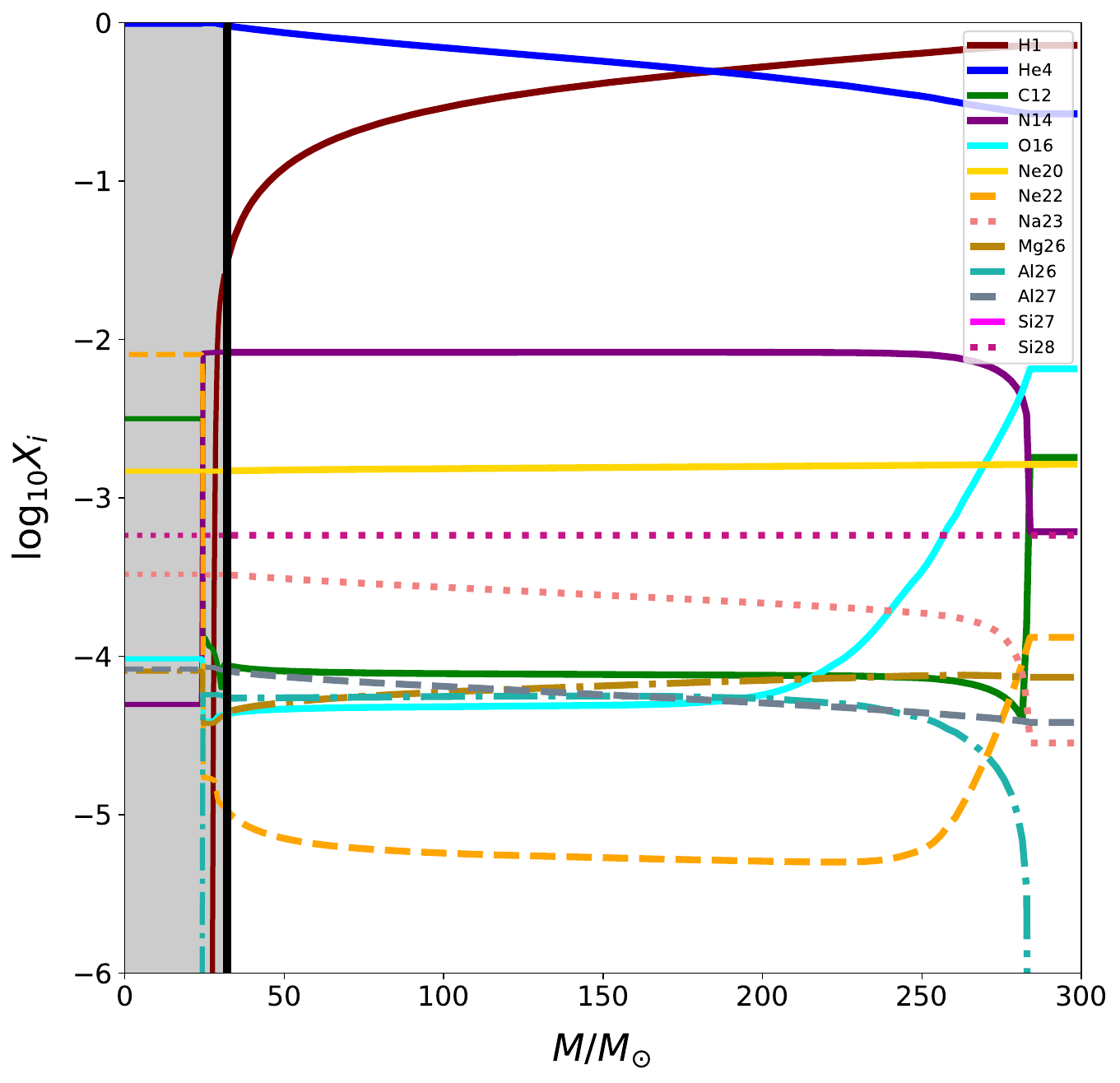}
    \caption{Time evolution of the surface composition during the MS coupled with the composition of the interior of the star (grey shaded area) at the end of the MS (both in mass fraction units and using a log-scale) for a model with an initial mass of 300\Mdot\ and the V11 wind prescription (where the legend details the various isotopes with coloured solid/dashed lines). Given that mass loss reduces the total mass of the star with time, the time evolution goes from right to left and the material ejected during the MS corresponds to the white area (right, non-shaded), with the hydrogen-exhausted stellar interior shown by the grey shaded region (left). The black solid line illustrates the TAMS where the surface evolution at core H-exhaustion occurs, and the central abundances are then shown in the grey region at the same TAMS point.}
    \label{fig:300S22TAMS}
\end{figure}
\begin{figure}
    \centering
    \includegraphics[width = \columnwidth]{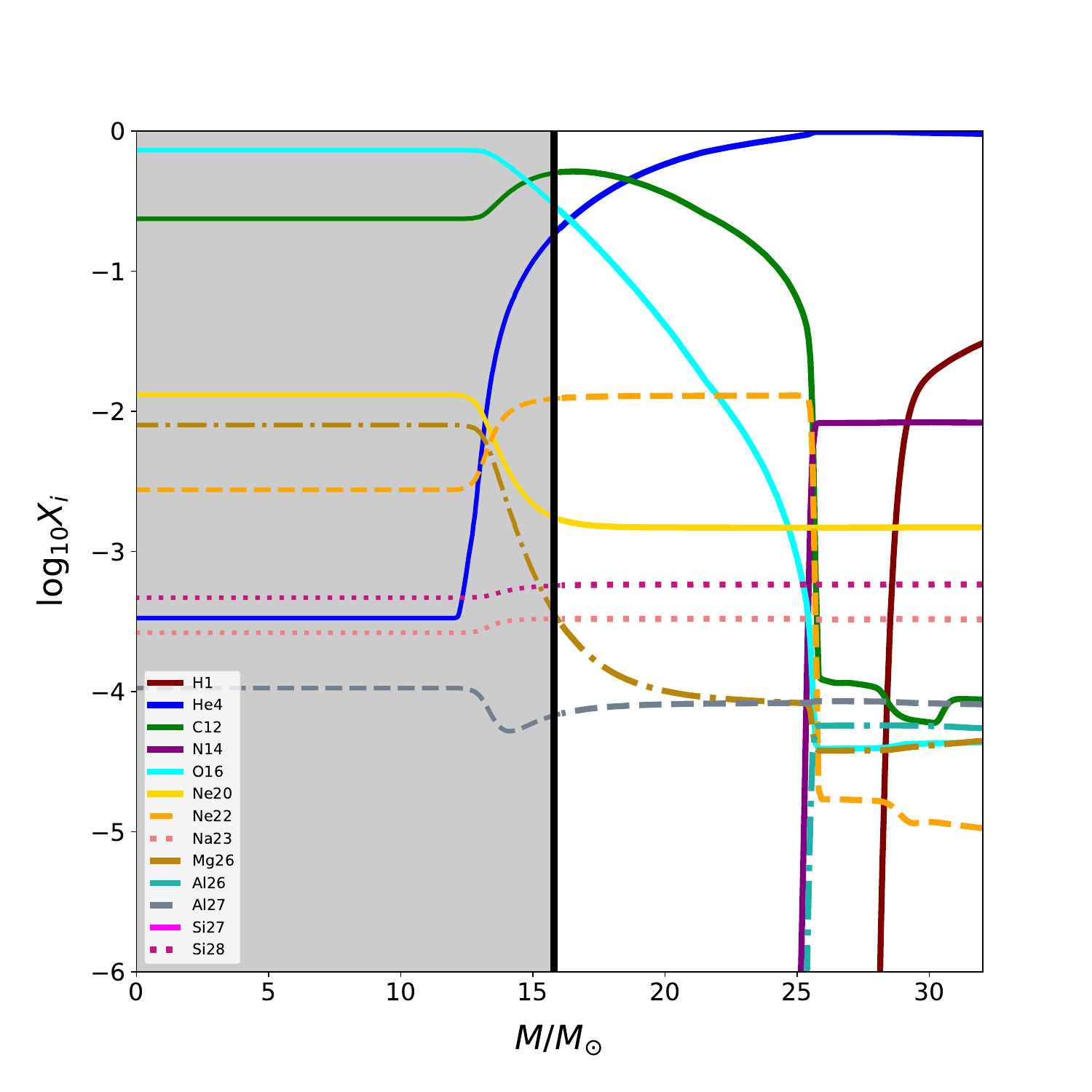}
    \caption{Time evolution of the surface composition during core He-burning in log-scale as a function of stellar mass with the interior composition shown at the end of core He-burning, for a model with an initial mass of 300\Mdot\ applying the V11 wind prescription. The He-exhausted core is shown in the grey shaded region (left) while the ejected material lost during the core He-burning phase can be seen in white (right).}
    \label{fig:300HeTAMS}
\end{figure}

Our implementation of V11 winds follows that of \cite{Sabh22} such that the V01 and V11 prescriptions are connected at the observed transition point and the maximum mass-loss rate of the 2 recipes (as a function of L, M, T, Z, and $\varv_\infty$) is adopted at each time-step. Therefore stars which are near the transition point, or evolve beyond the transition point adopt the appropriate mass-loss rate at all evolutionary stages. In our model grid, all VMS (M$>$100\Mdot) lie beyond the transition point and as such apply the V11 prescription as shown in equation \ref{V11} throughout their evolution. On the other hand, the lower mass model calculated with M$_{\rm{init}}$ $=$ 50\Mdot\ evolves below the transition point and applies the V01 prescription outlined in equation \ref{V01} throughout its entire evolution. Finally, the 80\Mdot\ model already begins its evolution near the transition point (at \Zdot, M$_{\rm{trans}}$ $\sim$ 76\Mdot) and as a result switches between V01 and V11 dependencies in line with the wind physics discussed in \cite{vink11} and \cite{Sabh22}.

\begin{table*}
    \centering
    \begin{tabular}{|c|c|c|c|c|c|c|c|c|c|}
    \hline
    $M_{\rm{i}}/\rm{M}_{\odot}$ & $M_{\rm{TAMS}}/\rm{M}_{\odot}$ &  $M_{\rm{\alpha}}/\rm{M}_{\odot}$ &  $M_{\rm{He-TAMS}}/\rm{M}_{\odot}$ &  $M_{\rm{CO}}/\rm{M}_{\odot}$ &  $M_{\rm{f}}/\rm{M}_{\odot}$ & $\tau_{\rm{MS}}/\rm{Myr}$ & $\tau_{\rm{He}}/\rm{Myr}$ & $\tau_{\rm{CO}}/\rm{yr}$ &$\dot{M}$ (Dutch, V01, V11) \\
    \hline \hline 

       100  & 32.95 & 29.57 & 16.20 & 15.82 & 15.94 & 3.12 & 0.379 & 6044 & V11 \\
       200  & 31.82 & 29.57 & 16.16 & 15.27 & 15.72 & 2.76 & 0.381 & 6095 & V11 \\
       300  & 32.20 & 28.71 & 16.05 & 15.47 & 15.80 & 2.62 & 0.381 & 6083 & V11 \\
       400  & 32.46 & 28.71 & 16.03 & 14.49 & 15.85 & 2.54 & 0.380 & 6065 & V11 \\       
       500  & 32.66 & 29.25 & 16.15 & 14.53 & 15.89 & 2.48 & 0.380 & 6066 & V11 \\      
       \hline
       100  & 61.86 & 46.82 & - & - & - & 3.05 &-&-& V01 \\
       200  & 98.37 & 41.67 & - & - & - & 2.46&-&-& V01 \\
       300  & 132.20 & 125.846 & - & - & - & 2.25&-&-& V01 \\
       400  & 161.33 & 156.636 & - & - & - & 2.14&-&-& V01 \\
       500  & 189.33 & 185.432 & - & - & - & 2.07&-&-& V01 \\
       \hline
    \end{tabular}
    \caption{Key characteristics of the model grid with masses provided at the ZAMS (M$_{\rm{i}}$), at the end of core H-burning (M$_{\rm{TAMS}}$), end of core He-burning (M$_{\rm{He-TAMS}}$), and final masses (M$_{\rm{f}}$). We also provide core masses at the end of core H-burning (M$_{\rm{\alpha}}$), and at the end of core C-burning (M$_{\rm{CO}}$).  Evolutionary timescales are provided for the MS ($\tau_{\rm{MS}}$), core He-burning phase and core CO-burning phase. The mass-loss prescription applied to the model is provided in the final column (V01,V11), as outlined in Sect.\ref{sectmassloss}. Note that the V01 models are calculated during the core H-burning phase only.}
    \label{tab:grid}
\end{table*}
\section{Results}\label{sectresults}
In this section, we present the initial results of our model grid for VMS masses ranging from 100-500\Mdot. In particular we focus on the nucleosynthesis and key characteristics of VMS evolution, resulting from two sets of models with one set implementing enhanced, optically thick, V11 stellar winds, and the other including the standard, optically thin, V01 winds, as outlined in Sect. \ref{method}.
\subsection{Effect of stellar winds on VMS}
Initially, we compare the effect of applying the enhanced V11 wind prescription in line with observations of VMS and theoretical predictions of enhanced mass loss above the transition point, with the standard O star winds of V01 in Fig. \ref{fig:V01V11compare}. We show the time evolution of surface composition ($^{1}$H, $^{4}$He, $^{12}$C, $^{14}$N and $^{16}$O) for a 300\Mdot\ star, as a function of total stellar mass, where mass is lost with time from right to left. By comparing the grey shaded region which illustrates the TAMS mass for each model, we can see that the wind yields and post-MS evolution are significantly affected by the MS mass-loss rate. Figure \ref{fig:V01V11compare} highlights the primary difference for VMS winds where the final TAMS mass differs by 100\Mdot. The subsequent effect of the additional mass lost by the V11 model during this MS phase is evident by the increased amounts of He and N ejected (white region of Fig.\ref{fig:V01V11compare} (b) when compared to the larger white region in (a)).  We explore this further for a wider range of isotopes in Figs. \ref{fig:300S22TAMS} and \ref{fig:300V01TAMS}, and discuss the key effects in Sect. \ref{sectMS}. We present an overview of the TAMS masses and evolutionary timescales of our models in Table \ref{tab:grid}.

\subsection{Nucleosynthesis}\label{sectnucleo}
VMS are extremely efficient nuclear fusion generators, with convective cores which are a significant fraction of their total mass, and burning timescales of just a few Myrs. Stars with an initial mass greater than $\sim$ 8\Mdot\ fuse H into He via the CNO-cycle as opposed to the p+p chain due to their increased central temperatures and relatively lower central densities. The heavier elements in the CNO-cycle act as catalysts with the net result of converting H into He. Initially the CN-cycle processes $^{12}$C towards $^{15}$N by proton-capture before returning to $^{12}$C again via (p,$\alpha$) reactions. This CN-cycle reaches equilibrium in $\sim$ 10$^{4}$ yr with a factor of 10 increase in $^{14}$N. Simultaneously, the second CNO-cycle converts $^{15}$N into $^{16}$O - $^{17}$F - $^{17}$O before returning to $^{14}$N again by (p,$\alpha$). Therefore, as a result of the CNO-processing during core H-burning in massive stars, the initial C and O abundances are depleted at the expense of producing $^{14}$N. 

At sufficiently high temperatures (T$_{c}$ $\sim$ 5 $\times$ 10$^{7}$ K), secondary reactions can occur during core H-burning. This includes the NeNa-cycle which processes the $^{20}$Ne into $^{22}$Ne and $^{23}$Na before returning to $^{20}$Ne again. The processed material can result in an observable increase in $^{23}$Na.

In addition, at T $>$ 10$^{7}$ K massive stars produce $^{26}$Al via the MgAl-cycle where $^{24}$Mg is converted to $^{25}$Al - $^{25}$Mg - $^{26}$Al before decaying to $^{26}$Mg or proton-captures to $^{27}$Si. Interestingly, the ground state $\tau_{1/2}$ of $^{26}$Al,
\citep[which makes up $\sim$ 77\% of $^{26}$Al synthesised;][]{laird23} is $\sim$ 0.7 Myr allowing observations to trace the production of $^{26}$Al as it decays. In fact, COMPTEL has observed $\sim$ 3\Mdot\ of radioactive $^{26}$Al in the Galactic plane of the Milky Way \citep{diehl95}, where massive star-forming regions are present. Therefore, massive star nucleosynthesis is expected to be crucial for explaining the presence of such an abundance of $^{26}$Al in our Galaxy \citep{laird23}. We note that in our MESA models, the nuclear reaction network combines the ground and isomeric state of $^{26}$Al, such that our $^{26}$Al wind yields could be reduced by approximately 23\% for $\gamma$-ray observation comparisons, as the isomeric component will decay to $^{26}$Mg effectively instantaneously ($\tau_{1/2}$ $=$ 6.35 s). We refer to a forthcoming future work for updated $^{26}$Al reaction rates and independent ground and isomeric states of $^{26}$Al for the precise yields produced of this isotope.

Figure \ref{fig:300S22TAMS} presents the nucleosynthesised material from core H-burning which would be lost in stellar winds via the V11 prescription. Isotopes are shown in logarithmic-scale and represented as a function of stellar mass. In this case we display the abundances of a 300\Mdot\ star evolving at solar Z. Each isotope evolves from right to left as the star synthesises material and loses mass through stellar winds on the MS. The ejected mass lost during core H-burning can be seen in white (right), with the H-exhausted stellar interior shown by the grey shaded region (left). As the star loses mass during the core H-burning phase (about 90\% of the star's entire lifetime), the total mass is reduced significantly from 300\Mdot\ to $\sim$30\Mdot. This is due to strong stellar winds experienced by the most massive stars which evolve close to the Eddington limit. With such large convective cores, these VMS are almost fully mixed, leading to nuclear fusion-products, like N, being exposed at the stellar surface early in the evolution. With strong outflows stripping these outer layers, the contribution of VMS winds on their environment is significant. Therefore, the white region showcases the stellar wind yields that are expected for each isotope (with the legend detailing the various isotopes with coloured solid/dashed lines). 

We can see from 220 $\lesssim$\,M /\Mdot$\, \lesssim$\, 280 that the C and O abundances are reduced at the expense of increased N due to the CNO-cycle. We also find an increase in $^{26}$Al during this phase as a result of the MgAl-cycle, as well as increased $^{23}$Na and reduced $^{22}$Ne due to the NeNa-cycle. Interestingly, the crossover from H to He enhancement seen at M$\approx$200\Mdot\ represents the chemically-homogeneous nature of these VMS which display surface H abundances which can be used as a `clock' to infer their core's evolutionary stage \citep{higgins22}. At this point, the central abundance is already exposed at the stellar surface, meaning the outer H-rich layers have been stripped from the star. The increase by a factor of 10 in $^{14}$N at $\approx$ 280\Mdot\ showcases CN-equilibrium which is reached quickly on the MS, with a comparable increase in $^{23}$Na. The abundance of $^{20}$Ne remains constant relative to the initial composition due to the regeneration of $^{20}$Ne at the end of the NeNa-cycle.

The post-MS is displayed in Fig. \ref{fig:300HeTAMS} where He-processed material is displayed in the white region (right) leaving the He-exhausted core in the grey-shaded region (left). As in Fig.\ref{fig:300S22TAMS}, the evolution of various isotopes goes from right to left as the star loses mass through stellar winds, and the elements shown in white will be lost in these winds while the core in grey retains any further processed material. We present the continuation of the 300\Mdot\ model showcased in Fig.\ref{fig:300S22TAMS}, for the core He-burning stage of evolution, which also implements the enhanced VMS wind of V11. In Fig. \ref{fig:300HeTAMS}, we do not include the already lost MS wind matter, but only include wind yields during the He-burning phase in white, allowing for a more detailed study of elements processed during core He-burning. We note that while we present a 300\Mdot\ model for the post-MS, all models which include the V11 wind result in the same final mass and element structure. Therefore, our 100\Mdot\ model could be discussed interchangeably here (see Fig. \ref{fig:yield100HeTams}).

We find that at the onset of core He-burning (M$\approx$ 26\Mdot), the N-rich material from the MS is quickly reprocessed into $^{22}$Ne which is enriched by a factor of 1000. This would make $^{22}$Ne a strong spectroscopic observable of early He-burning nucleosynthesis in stripped stars, particularly as it is $\sim$10 times more abundant than $^{20}$Ne. We also see that as He converts into C and O. Their abundances increase by a factor of 100 and 1000 respectively (M$\approx$ 25\Mdot). We note an increase in $^{26}$Mg at the expense of $^{26}$Al at the same point.  This demonstrates that classical WR stars do not eject meaningful amounts of $^{26}$Al before it has decayed to $^{26}$Mg, due to the lack of H remaining.
\begin{figure}
    \centering
    \includegraphics[width = \columnwidth]{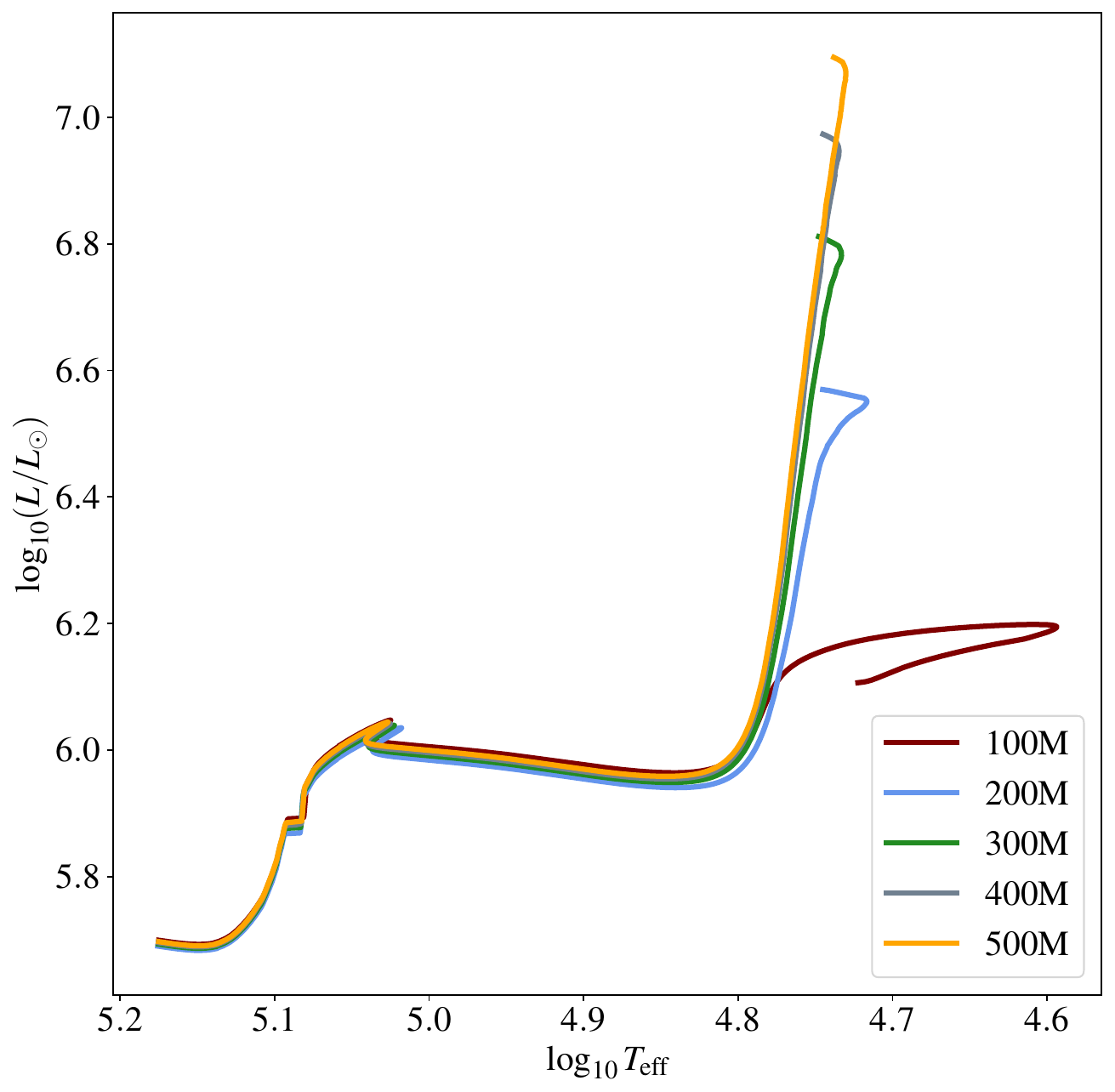}
    \caption{Hertzsprung-Russell diagram of the grid of models comprising initial masses of 100\Mdot, 200\Mdot, 300\Mdot, 400\Mdot\ and 500\Mdot, including the enhanced wind prescription V11. }
    \label{fig:HRDV11}
\end{figure}
\begin{figure}
    \centering
    \includegraphics[width = \columnwidth]{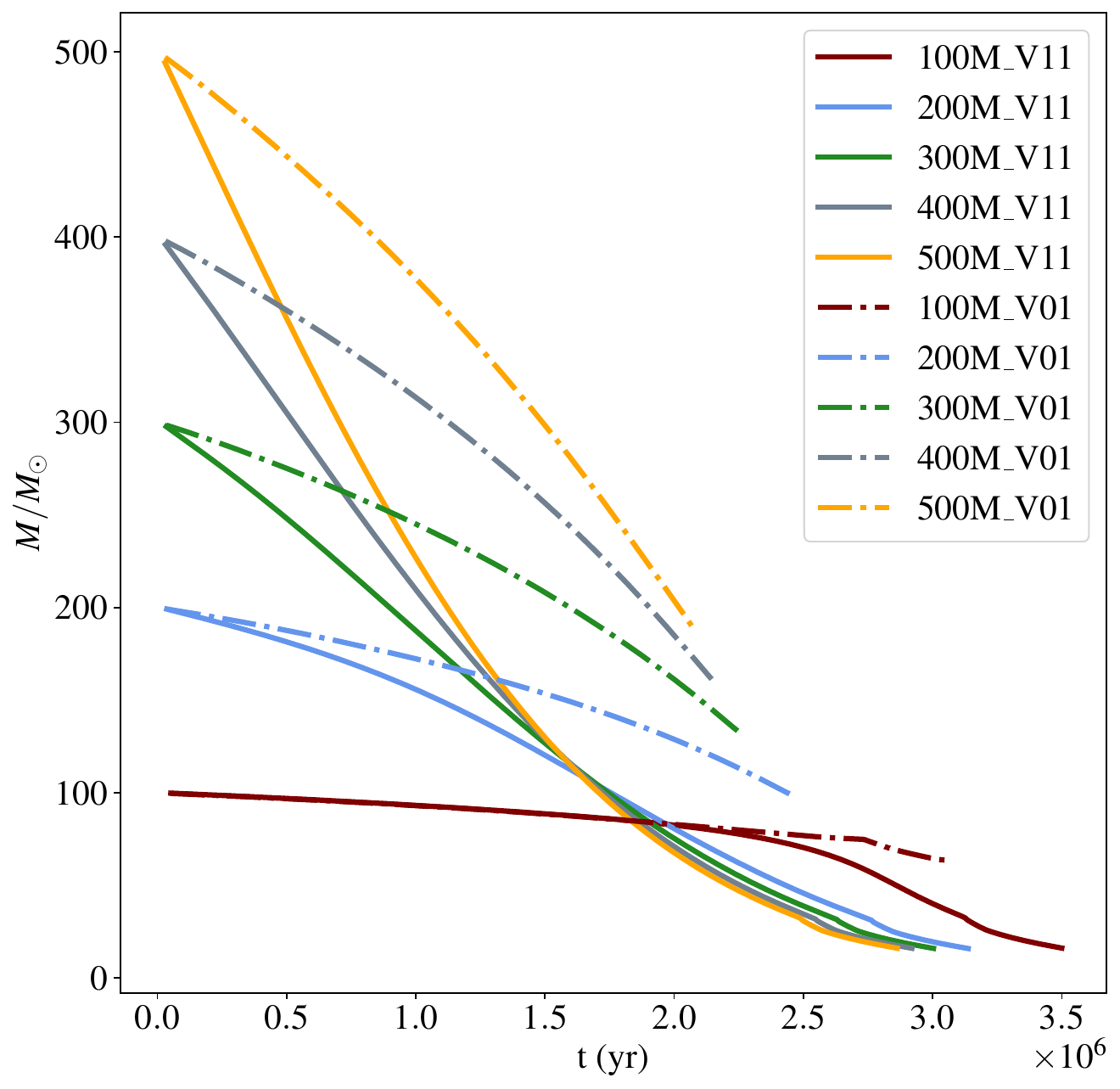}
    \caption{Mass evolution of the grid of models comprising initial masses of 100\Mdot, 200\Mdot, 300\Mdot, 400\Mdot\ and 500\Mdot, including the wind prescriptions V11 (solid) and V01 (dashed) during core H-burning only. }
    \label{fig:MtS22V01}
\end{figure}
\subsection{Evolution of VMS}\label{sectevolution}
We evolve VMS models with initial masses ranging from 100-500\Mdot\ adopting the appropriate enhanced-wind prescription from \cite{Sabh22} for stars above the transition point. This $\Gamma$-dependent mass loss results in a self-regulatory effect where stars lose a significant fraction of their total mass on the MS leading to a drop in luminosity at a constant effective temperature as has been observed in the Arches cluster of the Milky Way and 30Dor in the LMC. We extend this $\Gamma$-dependent mass loss through the post-MS stages of evolution in V11 models as the relative dependencies are consistent with that of WR stars \citep{Sander20}. This results in a second drop in luminosity after the onset of core He-burning, see Fig. \ref{fig:HRDV11}. Comparisons by Sabhahit et al. (2023) have shown that the absolute rates of the hydrodynamically-consistent WR rates by \cite{Sander20} are in good agreement with the extension of the V11 rates during core He-burning. The 100\Mdot\ model initially evolves to cooler effective temperatures before losing sufficient mass to evolve quasi-chemically homogeneously, while stars with an initial mass greater than 200\Mdot\ already begin the MS and evolve chemically homogeneously.

The enhanced wind of V11 models (solid lines) results in a steep drop in mass with a mass turnover point at 1.6\,Myrs as previously explored in \cite{higgins22}. We compare the mass evolution of V11 models and V01 models during core H-burning in Fig. \ref{fig:MtS22V01}, showcasing the consequences of the weaker wind rate of V01 designed for O stars below the transition point. We find that the enhanced wind models converge to a TAMS mass of $\sim$ 32\Mdot\ while models applying the V01 wind have TAMS masses ranging from 60\Mdot\ up to 190\Mdot. The net yields and ejected mass lost on the MS are impacted significantly by this change in mass. This effect can also be seen by comparing Fig. \ref{fig:300S22TAMS} with the reduced winds in Fig. \ref{fig:300V01TAMS}. The TAMS mass in Fig. \ref{fig:300V01TAMS} is 132\Mdot\ reducing the contribution of wind yields substantially.

\begin{table*}
    \centering
    \begin{tabular}{c|c|c|c|c|c|c|c|c|c|c|c|c|c|c|c}
    \hline
        $M_{\rm{i}}$ &$Y_{\rm{s}}$ &N/C & N/O & Ne/He & O/He & C/N & C/He \\
        \hline \hline
&&&&H-exhaustion	\\	
\hline
50	&2.660E-1	&3.393E-1	&9.369E-2	&4.961E-4	&2.458E-2	&2.947 & 6.788E-3\\
80	&8.047E-1	&1.108E+2	&1.683E+2	&6.462E-6	&6.146E-5	&9.022E-3& 9.330E-5\\
100	&9.572E-1	&9.403E+1	&1.921E+2	&1.150E-5	&4.518E-5	&1.063E-2& 9.229E-5\\
200	&9.532E-1	&9.526E+1	&1.899E+2	&1.076E-5	&4.591E-5	&1.050E-2& 9.150E-5\\
300	&9.548E-1	&9.479E+1	&1.908E+2	&1.099E-5	&4.561E-5	&1.055E-2& 9.179E-5\\
400	&9.557E-1	&9.451E+1	&1.913E+2	&1.114E-5	&4.543E-5	&1.058E-2& 9.197E-5\\
500	&9.564E-1	&9.429E+1	&1.917E+2	&1.125E-5	&4.530E-5	&1.061E-2& 9.211E-5\\
      \hline						
&&&&He-exhaustion						\\
\hline

50	&5.095E-1	&9.473	&2.427	&9.178E-5	&4.596E-3	&1.056E-1&1.177E-3 \\
80	&1.544E-1	&$\sim$0	&$\sim$0	&7.725E-2	&2.270	&8.117E+13& 3.083\\
100	&1.980E-1	&$\sim$0	&$\sim$0	&6.258E-2	&1.399	&1.067E+14&2.559 \\
200	&2.079E-1	&$\sim$0	&$\sim$0	&5.989E-2	&1.265	&1.147E+14& 2.457\\
300	&2.043E-1	&$\sim$0	&$\sim$0	&6.084E-2	&1.311	&1.120E+14&2.493 \\
400	&2.020E-1	&$\sim$0	&$\sim$0	&6.147E-2	&1.343	&1.103E+14& 2.517\\
500	&2.002E-1	&$\sim$0	&$\sim$0	&6.196E-2	&1.368	&1.092E+14&2.536 \\
	      \hline			
&&&&O-exhaustion						\\
\hline

50	&5.195E-1	&1.460E+1	&2.567	&8.903E-5	&4.412E-3	&6.848E-2& 7.757E-4\\
80	&1.369E-1	&$\sim$0	&$\sim$0	&8.569E-2	&2.781	&7.188E+13& 3.384\\
100	&1.761E-1	&$\sim$0	&$\sim$0	&6.964E-2	&1.741	&9.492E+13& 2.831\\
200	&1.852E-1	&$\sim$0	&$\sim$0	&6.667E-2	&1.578	&1.017E+14& 2.723\\
300	&1.818E-1	&$\sim$0	&$\sim$0	&6.774E-2	&1.636	&9.937E+13& 2.763\\
400	&1.797E-1	&$\sim$0	&$\sim$0	&6.843E-2	&1.673	&9.794E+13& 2.788\\
500	&1.782E-1	&$\sim$0	&$\sim$0	&6.896E-2	&1.703	&9.694E+13& 2.807\\
      \hline
    \end{tabular}
    \caption{Relative surface abundances in mass fractions for a range of initial masses provided in solar mass units. The surface abundance ratios are provided from V11 models and are shown for three evolutionary stages. }
    \label{tab:observedratios}
\end{table*}

\subsection{Observable surface chemical signatures}\label{observation}
The surface evolution of various isotopes shown in Figs. \ref{fig:300S22TAMS} and \ref{fig:300HeTAMS} illustrates the dominant isotopes through each evolutionary stage. However, we can also showcase the change in surface enrichment by providing relative surface abundances at particular evolutionary times. Therefore, we consider surface abundances at core H-exhaustion, He-exhaustion and O-exhaustion which represents the surface properties at the black line of Figs. \ref{fig:300S22TAMS} and \ref{fig:300HeTAMS}, as well as the final surface profile of each model.

We provide surface ratios of $^{4}$He, $^{12}$C, $^{14}$N, $^{16}$O, and $^{22}$Ne in mass fractions in Table \ref{tab:observedratios} to compare with abundance ratios which may be observed in H-burning WNh stars or post-MS WR stars. As in \cite{Ekstroem2012}, we present ratios of N/C and N/O abundances, as well as surface $^{4}$He, finding that our VMS results correspond to the same order of magnitude as their M $>$ 60\Mdot\ models. We compare our 80-500\Mdot\ model ratios of surface N/C to the observed VMS in the Arches cluster ($\sim$\Zdot), and find an excellent agreement between our H-exhaustion ratios (N/C $\sim$ 90-110) and the sample of WNh stars in \citet[Fig. 8; ][]{martins08} which are observed to have log N/C $\sim$ 2. We also provide O/He ratios for later evolutionary stages, comparable to \citet[Fig. 10; ][]{MM94}, and \cite{Crowther+2002}. The C/He and O/He ratios at He-exhaustion and O-exhaustion suggest that our 50\Mdot\ model best represents the WR stars from \cite{Crowther+2002}. 

We note that the N/C and N/O ratios will be substantially higher on the MS than the post-MS due to the CNO-cycle, while the Ne/He and O/He will increase in the post-MS as $^{14}$N is processed into $^{22}$Ne and $^{16}$O is produced during core He-burning. Interestingly, the surface $^{4}$He is almost 100\% at the end of core H-burning for VMS while the 50\Mdot\ model is not enriched in $^{4}$He at all during core H-burning. As previously mentioned, this is a result of 2 features of VMS evolution, the fully mixed interior where a large fraction of the star is occupied by the convective core, and strong stellar winds stripping the exterior envelope. A similar effect can be seen for the 50\Mdot\ ratios of N/C and N/O which increase in later burning stages due to delayed stripping of this earlier-processed material, compared to VMS which display surface abundances which are representative of their core abundances.
\begin{table*}
    \centering
    \begin{tabular}{c|c|c|c|c|c|c|c|c|c|c|c|c|c|c|c}
    \hline
        $M_{\rm{i}}/\rm{M}_{\odot}$ &$\dot{M}$ & $^{1}$H & $^{4}$He & $^{12}$C & $^{14}$N & $^{16}$O & $^{20}$Ne & $^{22}$Ne & $^{23}$Na &$^{26}$Mg &$^{26}$Al &$^{27}$Al & $^{28}$Si \\
        \hline \hline
      100   & V11 & 27.418 & 38.339 & 0.033 & 0.422 & 0.118 & 0.105 & 0.002 & 0.013 & 4.88E-3 & 2.271E-3 & 3.454E-3 & 0.039\\
      200   & V11 & 67.603 & 97.182 & 0.041 & 1.234 & 0.148 & 0.261 & 0.003 & 0.038 & 0.011 & 7.391E-3 & 0.009 & 0.097\\
      300   & V11 & 115.623 & 146.385 & 0.045 & 2.046 & 0.167 & 0.416 & 0.004 & 0.060  &0.017 & 0.013 & 0.015 & 0.154\\
      400   & V11 & 166.382 & 192.622 & 0.050 & 2.846 & 0.196 & 0.571 & 0.005 & 0.082  &0.023 & 0.018 & 0.021 & 0.221\\
      500   & V11 & 222.696 & 239.014 & 0.065 & 3.649 & 0.264 & 0.736 & 0.007 & 0.104  &0.029 & 0.023 & 0.026 & 0.272\\
     \hline
      100   &  V01 & 22.805 & 12.598 & 0.030 & 0.166 & 0.117 & 0.058 & 2.248E-3 & 4.339E-3 & 3.170E-3 & 4.005E-4 & 1.491E-3 & 0.021\\
      200   & V01 & 46.798 & 52.162 & 0.037 & 0.682 & 0.140 & 0.158 & 3.008E-3 & 0.020 & 7.549E-3 & 3.313E-3 & 5.328E-3 & 0.058\\
      300   & V01 & 68.392 & 94.944 & 0.040 & 1.221 & 0.150 & 0.258 & 3.795E-3 & 0.038 & 0.011 & 6.796E-3 & 9.865E-3 & 0.096\\
      400   & V01 & 91.733 & 140.717 & 0.045 & 1.794 & 0.164 & 0.366 & 4.867E-3 & 0.056 & 0.014 & 0.011 & 0.015 & 0.137\\
      500   & V01 & 119.558 & 186.017 & 0.056 & 2.380 & 0.197 & 0.480 & 6.361E-3 & 0.075 & 0.018 & 0.015 & 0.020 & 0.180\\
      \hline
    \end{tabular}
    \caption{Ejected masses calculated with equation \eqref{EMeq} for V11 and V01 models during core H-burning only. Initial masses and ejected masses are provided in solar mass units.}
    \label{tab:MSyieldcomparison}
\end{table*}

\section{Yields and ejected masses}\label{sectyield}
In this section we provide calculations of ejected masses and net wind yields for V11 models until core O-exhaustion (Table \ref{tab:EMyieldsV11}), as well as a comparison of MS ejected masses for both sets of models, applying V11 and V01 winds (Table \ref{tab:MSyieldcomparison}). We discuss the key variations in ejected isotopes when implementing enhanced VMS winds or O-star winds, with consequences for galactic chemical evolution (GCE). Since most of the ejecta are lost during core H-burning, the key differences in ejected (element) masses will occur as a result of the MS mass-loss prescription. 

We adapt the relations from \cite{hirschi05} for our yield calculations. The net stellar wind yield calculated for a star of initial mass, $m$, and isotope, $i$, is defined as:
\begin{equation}\label{yieldeq}
    m^{\rm wind}_{i} = \int_{0}^{\tau(m)} \dot{M}(m, t) \,[{X}^{S}_{i}(m, t) - {X}^0_{i}] \,dt 
\end{equation}
where $\dot{M}$ is the mass-loss rate, ${X}^{S}_{i}$ is the surface abundance of a given isotope, and ${X}^0_{i}$ is the initial abundance of a given isotope (see Table \ref{tab:abundances}), integrated from the ZAMS until $\tau(m)$, the final age of the star. 
We also calculate ejected masses (EM) of each isotope, $i$, by using:
\begin{equation}\label{EMeq}
    EM_{im} = \int_{0}^{\tau(m)} \dot{M} \, {X}^{S}_{i}(m, t) \,dt .
\end{equation}
We find that all V11 models reach O-exhaustion with the same final mass ($\sim$ 16\Mdot) and structure, and assume that they all collapse to form black holes without a supernova. Therefore, we implement the above wind yield equations such that the ejected masses and net yields are all attributed to stellar winds.

We present the complete table of ejected masses (top) and net yields (bottom) in solar mass units for our V11 model grid in Table \ref{tab:EMyieldsV11}. We find that with increased initial mass, more $^{1}$H, $^{4}$He and $^{14}$N are expelled, as would be expected. However, we find that the ejected masses of $^{12}$C, $^{16}$O, and $^{22}$Ne are relatively constant with initial mass since they are post-MS products. This demonstrates the dominant role that MS mass loss plays on the entire evolution of VMS, including their total yields. Furthermore, we can see from the increasing $^{4}$He ejecta that much of the element is lost in the MS before converting into $^{12}$C in the post-MS. Hence, the H-processed elements will produce the majority of stellar wind yields. Interestingly, we see an increase in the amount of $^{20}$Ne, $^{23}$Na, $^{26}$Al and $^{27}$Al ejected with higher initial masses, suggesting that the most massive stars may be responsible for polluting their environments with these trace elements. This is important for comparisons with $\gamma$-ray observations, and globular clusters which show enrichment of $^{23}$Na and $^{27}$Al \citep{bastianlardo18}.

The net wind yields, useful for GCE calculations, are provided in the lower section of Table \ref{tab:EMyieldsV11}. Negative values show that the net element mass has been processed into another element, for example $^{1}$H yields are negative at the expense of $^{4}$He. Similarly, for M$_{\rm{init}}$ $>$ 200\Mdot\ the additional NeNa and MgAl-cycles produce $^{23}$Na at the cost of $^{20}$Ne and $^{26}$Al at the expense of $^{24}$Mg. In fact, increased $^{20}$Ne abundances also demonstrates evidence of previously processed $^{23}$Na. Moreover, at these high initial masses the stars evolve chemically-homogeneously, so stars with M$_{\rm{init}}$ $>$ 200\Mdot\ show that $^{14}$N produces a net positive yield as a result of processing $^{16}$O during the CNO-cycle, while the 100\Mdot\ model which is not fully-mixed retains a positive yield in both elements. 

\begin{figure}
    \centering
    \includegraphics[width = \columnwidth]{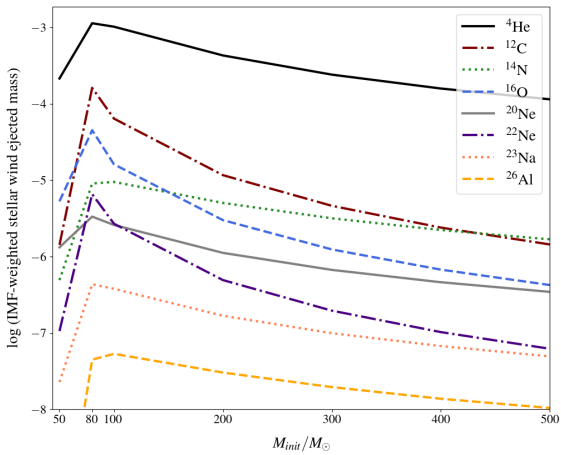}
    \caption{An IMF-weighted contribution of the logarithmic stellar wind ejected masses shown for models including the V11 wind prescription. The IMF included here adopts the \citet{salpeter} relation where $M^{-2.35}$.}
    \label{fig:salpeterIMF}
\end{figure}

We also provide an IMF-weighted contribution of our stellar wind yields in Fig. \ref{fig:salpeterIMF}. We adopt the relation from \citet{salpeter} for our VMS study, where $M^{-2.35}$. We compare the IMF-weighted ejected masses applied in Fig. \ref{fig:salpeterIMF} with a top-heavy IMF in Fig. \ref{fig:IMFschneider} finding similar results. Figure \ref{fig:Mini_yield} comparatively shows the stellar wind yields divided by initial mass as a function of the initial mass of models implementing the V11 wind prescription. This enables a direct comparison of the relative wind yield for each stellar mass. 

\subsection{Main sequence}\label{sectMS}
In Table \ref{tab:MSyieldcomparison} we explore the effects of MS mass loss on ejected masses of key isotopes. As previously discussed, the MS winds dominate the total yields, therefore we compare 2 sets of models with differing mass-loss rates on the MS (V11, top and V01, bottom). The ejected masses can easily be converted into the net wind yields, seen in Table \ref{tab:EMyieldsV11}, using the relation: 
\begin{equation}
     m^{\rm wind}_{i} = EM_{im} - {X}_{i}(M_{i} - M_{f}) 
\end{equation}
where the product of the mass lost and initial abundance, provided in Table \ref{tab:abundances}, is removed from the ejected mass of a given isotope. We only compare here the ejected masses of V11 and V01 models, since the mass is directly impacted by the wind prescriptions discussed.
\begin{table*}
    \centering
    \begin{tabular}{c|c|c|c|c|c|c|c|c|c|c|c|c|c|c}
    \hline
    $M_{\rm{i}}/\rm{M}_{\odot}$ & $^{1}$H & $^{4}$He & $^{12}$C & $^{14}$N & $^{16}$O & $^{20}$Ne & $^{22}$Ne & $^{23}$Na &$^{26}$Mg &$^{26}$Al &$^{27}$Al & $^{28}$Si \\
    \hline \hline 
      50   & 5.698 & 2.105 & 0.014 & 4.85E-3 & 0.052 & 0.013 & 1.04E-3 & 2.25E-4 & 5.85E-4 & 0 & 3.03E-4 &  4.60E-3\\
      80   & 22.131 & 33.552 & 4.788 & 0.268 & 1.337 & 0.099 & 0.195 & 0.0129 & 6.113E-3 & 1.33E-3 & 3.35E-3 &  0.036\\
      100   & 27.617 & 51.069 & 3.203 & 0.476 & 0.808 & 0.130 & 0.135 & 0.019 & 6.54E-3 & 2.68E-3 & 4.70E-3 & 0.049\\
      200   & 67.670 & 109.516 & 2.973 & 1.288 & 0.767 & 0.286 & 0.126 & 0.043 & 0.013 & 7.77E-3 & 0.011 &  0.107\\
      300   & 115.687 & 158.687 & 3.061 & 2.101 & 0.822 & 0.444 & 0.130 & 0.066 & 0.019 & 0.013 & 0.016 &  0.165\\
      400   & 169.343 & 206.126 & 3.125 & 2.903 & 0.880 & 0.602 & 0.134 & 0.088 & 0.025 & 0.018 & 0.022 &  0.223\\
      500   & 222.756 & 251.503 & 3.175 & 3.703 & 0.935 & 0.761 & 0.137 & 0.109 & 0.031 & 0.023 & 0.028 &  0.282\\
     \hline
      50   & -5.75E-5 & 5.99E-7 & -7.87E-7 & 3.64E-8 & -1.24E-10 & -2.55E-11 & -3.01E-11 & 2.89E-11 & -1.16E-12 & 0 & -5.99E-13 &  -9.08E-12\\
      80   & -22.932 & 16.906 & 4.675 & 0.230 & 0.928 & -3.56E-3 & 0.186 & 0.0111 & 1.49E-3 & 1.33E-3 & 9.47E-4 &  -4.69E-5\\
      100   &  -32.668 & 28.800 & 3.051 & 0.425 & 0.261 & -0.006 & 0.124 & 0.016 & 3.49E-4 & 2.68E-3 & 1.49E-3 &  -9.76E-6\\
      200   & -64.112 & 60.617 & 2.640 & 1.176 & -0.435 & -0.014 & 0.101 & 0.038 & -7.45E-4 & 7.77E-3 & 3.55E-3 &  1.06E-5\\
      300   & -87.268 & 83.717 & 2.548 & 1.927 & -1.036 & -0.019 & 0.092 & 0.057 & -1.96E-3 & 0.013 & 5.43E-3 & 2.27E-5\\
      400   & -107.453 & 103.880 & 2.431 & 2.667 & -1.634 & -0.025 & 0.083 & 0.077 & -3.31E-3 & 0.018 & 7.26E-3 & 3.36E-5\\
      500   & -126.185 & 122.608 & 2.300 & 3.406 & -2.234 & -0.030 & 0.073 & 0.095 & -4.72E-3 & 0.023 & 9.09E-3 &  4.40E-5\\
      \hline
    \end{tabular}
    \caption{Ejected masses (top) and net yields (bottom) calculated with equations \eqref{EMeq} and \eqref{yieldeq} respectively, for V11 models over the complete evolution until core O-exhaustion. Initial masses, ejected masses, and yields are provided in solar mass units.}
    \label{tab:EMyieldsV11}
\end{table*}
The most significant differences noted in Table \ref{tab:MSyieldcomparison} are of course $^{1}$H and $^{4}$He, where V11 models eject almost twice the amount of $^{4}$He compared to V01 models, see also Fig.\ref{fig:V01V11compare}. Moreover, the V11 models have higher ejected masses of $^{14}$N, due to the surface exposed fusion-products in V11 models. Interestingly, the ejected $^{26}$Al differs considerably with V11 producing up to 10 times more than V01 models. The difference in CNO masses highlights that V01 models do not reveal these core-processed materials early in the evolution. Moreover, the trace elements such as $^{20}$Ne, $^{23}$Na and $^{26}$Al are significantly impacted by the wind prescription with reduced ejected masses for V01 models. We note that the most significant impact on $^{26}$Al ejecta occurs at M$_{\rm{init}}$ $=$ 100\Mdot\ where V01 models predict a much lower value than V11 models. This suggests that previous studies of VMS could have under-predicted the contribution of VMS in the enrichment of $^{26}$Al as a result of wind-driving physics.


\subsection{Effect of VMS winds on the post-main sequence}\label{sectpostMS}

The post-MS evolution of VMS is severely impacted by the MS winds, dictating both the He-ZAMS mass and structure. From Fig.\ref{fig:V01V11compare} we can see that enhanced winds leave a much lower He-ZAMS mass than with standard winds. For models implementing the V11 wind (M$_{\rm{init}}$ $\geq$ 100\Mdot), all He-ZAMS masses are $\sim$ 32\Mdot\ with very similar chemical abundances, as can be seen by comparing the models with initial masses of 300\Mdot\ and 100\Mdot\ in Figs. \ref{fig:300HeTAMS} and \ref{fig:yield100HeTams} respectively.

The ejected masses and net wind yields are also completely dominated by the MS wind. By comparing Tables \ref{tab:MSyieldcomparison} and \ref{tab:EMyieldsV11} we see that only 0.021\Mdot\ of $^{20}$Ne, $\sim$0.05\Mdot\ of $^{14}$N and 0.0004\Mdot\ of $^{26}$Al is ejected in the post-MS for a 100\Mdot\ model. 
We find similar results for all initial masses which implement the V11 wind, showcasing that the MS wind dictates the total ejected masses and wind yields within $\sim$ 0.1-0.001\Mdot, or within 1-5\%. As a consequence of the MS wind deciding the He-ZAMS mass, by He-exhaustion the central and surface abundances, He-wind yields and ejected masses are all the same regardless of initial mass (see Figs. \ref{fig:yield100HeTams} and \ref{fig:300HeTAMS} for instance).

Interestingly, we find that the relative abundances of $^{12}$C and $^{22}$Ne increase dramatically during the core He-burning stage (see Fig. \ref{fig:300HeTAMS}, M$\sim$ 25\Mdot) and therefore, 10 times more are ejected during the post-MS. We highlight that models at core He-exhaustion show surface enrichment with a factor of 10 increase in $^{22}$Ne abundance compared to $^{20}$Ne, regardless of initial mass, suggesting that $^{22}$Ne could be the dominant Ne-isotope observed in WRs. This isotope may be a key tracer of WR evolution, suggesting that the impact of VMS on WR studies could be broader than previously expected. \cite{kob11} find that the solar $^{22}$Ne/$^{20}$Ne ratios are in good agreement with current GCE models for M $<$ 40\Mdot, which presents an interesting comparison to the VMS yields of $^{20}$Ne and classical WR yields of $^{22}$Ne.

From Fig.\ref{fig:V01V11compare} we can see, by comparing the MS wind effect on the He abundance that, for the same current mass, for example 150\Mdot, He-rich objects have stripped less of their envelope with low mass-loss rates (V01, Fig. \ref{fig:300V01TAMS}) than objects with the same current mass which have a lower surface He abundance with higher mass-loss rates (V11, Fig. \ref{fig:300S22TAMS}). Therefore MS stars with $Y=$0.8-1.0 suggest that either (i) VMS$\sim$300\Mdot\ stars do not exist, or (ii) V01 mass-loss rates are too low for these objects on the MS, since these surface abundances can act as a `clock' as described in \cite{higgins22}. Due to the large cores and strong winds, the surface evolution of H and He reveals the core evolution as well.

\section{Contribution of VMS compared to O-stars}\label{lowmass}
Canonical OB stars in the 8-20 \Mdot\ range likely end their lives in various supernovae, ejecting heavy elements into their host galaxy, leaving a compact remnant. The most massive O stars (30\Mdot $<$M$<$ 60\Mdot) eject material during their lives through stellar winds, though mostly in the form of $^{1}$H and $^{4}$He, with 10$^{-2}$ to 10$^{-6}$ \Mdot\ of heavier elements like $^{12}$C, $^{14}$N, and $^{16}$O \citep{hirschi05}. VMS can eject substantially higher masses of nuclear-processed elements, not only due to their enhanced winds, but as a result of stripping their outer envelope early on the MS, and their CHE nature, even expose the nuclear-burning core at the surface leading to increased net yields of elements such as $^{26}$Al, $^{14}$N, $^{20}$Ne and $^{23}$Na. With such high initial masses and strong winds, VMS can dominate the yields of an entire IMF in their host galaxy or cluster. 
\begin{table*}
    \centering
    \begin{tabular}{c|c|c|c|c|c|c|c|c|c|c|c|c|c|c|c|c}
    \hline
    $M_{\rm{i}}/\rm{M}_{\odot}$  & $^{4}$He & $^{12}$C & $^{14}$N & $^{16}$O & $^{20}$Ne & $^{22}$Ne &$^{23}$Na & $^{26}$Al\\
    \hline \hline 

50&	3.54E-10&	-4.66E-10&	2.15E-11&	-7.33E-14&	-1.51E-14&	-1.78E-14& 1.71E-14 & 0\\
80&	4.09E-03&	1.13E-03&	5.57E-05&	2.25E-04&	-8.63E-07&	4.50E-05 & 2.69E-06 & 3.22E-07\\
100&	4.56E-03&	4.84E-04&	6.74E-05&	4.14E-05&	-9.51E-07&	1.97E-05 & 2.54E-06 & 4.25E-07\\
200	&2.57E-03&	1.12E-04&	4.99E-05&	-1.85E-05	&-5.95E-07&	4.29E-06 & 1.61E-06	&3.30E-07\\
300	&1.65E-03&	5.01E-05&	3.79E-05&	-2.04E-05&	-3.73E-07	&1.81E-06 & 1.12E-06 &2.56E-07\\
400	&1.18E-03&	2.77E-05&	3.03E-05&	-1.86E-05&	-2.84E-07&	9.44E-07 & 8.76E-07	&2.05E-07\\
500	&9.13E-04&	1.71E-05&	2.54E-05&	-1.66E-05&	-2.23E-07&	5.44E-07  & 7.07E-07& 1.71E-07\\
\hline
50	&	1.25E-03&	8.28E-06&	2.87E-06&	3.08E-05&	7.69E-06&	6.18E-07& 1.33E-07	&0\\
80	&	8.13E-03&	1.16E-03&	6.49E-05&	3.24E-04&	2.40E-05&	4.72E-05& 3.12E-06	& 3.22E-07\\
100	&	8.09E-03&	5.08E-04&	7.54E-05&	1.28E-04&	2.06E-05&	2.14E-05 & 3.01E-06	& 4.25E-07\\
200	&	4.65E-03&	1.26E-04&	5.47E-05&	3.26E-05&	1.21E-05&	5.35E-06 & 1.83E-06	&3.30E-07\\
300	&	3.13E-03&	6.02E-05&	4.13E-05&	1.62E-05&	8.73E-06&	2.56E-06& 1.30E-06	& 2.56E-07\\
400	&	2.35E-03&	3.56E-05&	3.30E-05&	1.00E-05&	6.85E-06&	1.52E-06 & 1.00E-06	& 2.05E-07\\
500	&	1.87E-03&	2.36E-05&	2.76E-05&	6.96E-06&	5.67E-06&	1.02E-06 & 8.12E-07 &	1.71E-07\\   
\hline

    \end{tabular}
    \caption{IMF-weighted net yields (top) and ejected masses (bottom) calculated with equation \eqref{imfeq}, for V11 models over the complete evolution until core O-exhaustion. We adopt the IMF of \citet{Schneider18} where M$^{-1.90}$.}
    \label{tab:IMFEMV11H05}
\end{table*}

Table \ref{tab:EMyieldsV11} includes the ejected masses and wind yields of 50\Mdot\ and 80\Mdot\ stars, demonstrating the magnitude of VMS ejecta compared to O stars. The 50\Mdot\ model ejects $\sim$100 times less of each isotope ($^{12}$C, $^{14}$N, $^{16}$O, $^{22}$Ne, $^{23}$Na and $^{28}$Si) than the 80\Mdot\ and 100\Mdot\ models. This means that VMS of $\sim$100\Mdot\ would produce the same wind yields as 100 massive O-stars. Crucially, the wind contribution of $^{26}$Al is zero for a 50\Mdot\ star, suggesting that the wind contribution of the observed $^{26}$Al in the Galaxy is dominated by VMS.
Table \ref{tab:appV11MS} shows that the net yields of a 50\Mdot\ star are also negligible or negative as opposed to the VMS, relative to their initial abundances, indicating that O stars do not replenish their host galaxies in the way that VMS do.

We reiterate that the 50\Mdot\ model would represent a standard O star, while the 80-100\Mdot\ models are in the transition region and show properties of both O stars and VMS. While the 80-100\Mdot\ models are not fully mixed (CHE) like that of models with M$\geq$ 200\Mdot, they do experience enhanced winds. As a result, the 80\Mdot\ and 100\Mdot\ stars eject similar amounts of each isotope (on the same order as VMS). Interestingly, these models eject more $^{12}$C and $^{16}$O than their more massive counterparts. This is because the stars have lost less mass on the MS, and do not expose their cores during core H-burning, leaving increased amounts of $^{4}$He to be processed into $^{12}$C and $^{16}$O during the post-MS (now as stripped stars), where large amounts of these elements are then ejected as they are produced.

In contrast to previous work by \cite{martinet22} and \cite{brinkman19}, we find that the most massive stars are responsible for ejecting the primary contribution of $^{26}$Al due to the implementation of enhanced winds. Where previous studies have suggested that evolved, WR stars are responsible for the enrichment of $^{26}$Al, we find that the post-MS stage produces only $\lesssim$ 5\% of the total wind yields, and therefore, WR winds are not the dominant polluters of $^{26}$Al to their host environments. In fact, \cite{martinet22} adopt the higher mass-loss rates of \cite{NugisLamers}, designed for stripped WR stars, in their VMS models when the surface H abundance falls below 40\%, leading to higher $^{26}$Al yields than would be expected for the MS-winds of O stars, \citep{Vink01}.
We compare our enhanced wind models with the VMS models of \cite{martinet22} finding that our 80\Mdot\ model yields a factor of 10 more $^{26}$Al than the 85\Mdot\ model from \cite{martinet22}, while our 200\Mdot\ model ejects 15 times more $^{26}$Al than their comparable 180\Mdot\ model. Finally, we compare directly with their 300\Mdot\ model finding that our enhanced wind models yield 4 times more when compared with the implementation of WR winds on the MS.
Interestingly, while binary stellar models have been suggested to eject more $^{26}$Al due to stripping of enriched material, our single 80\Mdot\ model still ejects 10 times more than the 80\Mdot\ primary component from \cite{brinkman19,brinkman21}. These comparisons prove key in reproducing the $^{26}$Al-rich material in our Galaxy, and that the most appropriate wind physics is required to provide accurate constraints on the chemical yields from the most massive stars. While our models do not treat the ground and isomeric state of $^{26}$Al separately, the branching ratio of these states is approximately 77\% $^{26}$Al$_{\rm{g}}$ and 23\% $^{26}$Al$_{\rm{*}}$ in massive stars, therefore we estimate that our $^{26}$Al yields are upper limits, with a potential uncertainty which is approximately 23\%, in contrast to the orders of magnitude increase in $^{26}$Al yields from VMS with enhanced winds. This inconsequential uncertainty does not significantly change the overabundance of $^{26}$Al ejected by VMS, and in comparison to lower mass stars and other works which eject factors of 10 less, the relative uncertainty in our treatment of the ground and isomeric state does not impact our conclusion that VMS winds are the primary donors responsible for $^{26}$Al-enrichment in the Galaxy \citep{vinkbook}.

We note that while VMS likely form black holes at the end of their lives and as such do not produce supernova yields, for lower mass OB stars $\sim$ 15\Mdot\ a supernova ejecta may also be included in the yield calculation. In \cite{hirschi05} supernova yields and wind yields are both included, providing a complete overview of ejected isotopes in this mass range. Due to the supernovae contribution, lower mass (12-25\Mdot) models yield significant amounts of $^{4}$He, $^{12}$C and $^{16}$O with IMF-weighted values ranging from 10$^{-2}$ to 10$^{-3}$. We find that in comparison to our wind yields, the isotopes dominated by lower mass supernovae progenitors would be $^{12}$C, $^{16}$O and $^{20}$Ne. GCE models for a range of metallicities from \cite{kobkar20}, accounting for SNe only, can reproduce the $^{16}$O, $^{24}$Mg, and heavier elements of observations from galaxies of varied Z. But some elements are overproduced, mainly the second and third s-process peaks, which depends on the remaining C abundance prior to the SNe. They find that C, N and $\alpha$-elements may be ejected prior to collapse for massive stars to best reproduce the galactic evolution trend as observed. 

Similarly, \cite{limongi18} present IMF-weighted yields for \Zdot\ including the wind and supernovae contribution of rotating models, where the 13-25\Mdot\ models include supernovae yields while higher mass models only include wind yields. Their IMF-weighted total yields show that $^{12}$C, $^{16}$O and $^{20}$Ne are in good agreement with observations, suggesting that SNe dominate due to the IMF, and therefore the production of specific isotopes \citep[see also][]{nomoto13}. However, $^{14}$N and $^{26}$Al are under-produced, and some heavy isotopes are overestimated (for example Ga, Ge, As, and Rb). The total (wind and supernova) $^{12}$C yields of a 20\Mdot\ from \cite{limongi03,limongi18} are in line with our 50\Mdot\ wind yields, but are a factor of 10 lower than our VMS wind yields, showcasing that individual VMS will eject significantly more enriched material, though due to their scarcity will not dominate the $^{12}$C production of an entire population.

We investigate the contribution of O stars and VMS on the net yields and ejected masses of their host galaxy, applying an IMF from \cite{salpeter}. We compare with a top-heavy IMF from \cite{Schneider18} which was found for the 30Dor region of the LMC. Since we calculate models of VMS in this work, we compare with this IMF relation in order to test the effect of VMS on IMF-weighted ejected masses, however we note that by comparing Figs. \ref{fig:salpeterIMF} and \ref{fig:IMFschneider}, we find little difference in the IMF-weighted yields. We do not infer an upper mass in our IMF as in \cite{maeder92} or \cite{ritter18} since the `effective' upper mass limit as explored by \cite{vink18} relies on a number of uncertain properties such as the pre-MS accretion rate and mass-loss rate, the ignition of the H core as a function of the star formation process, and relative to each of these properties - the host Z content. 

Table \ref{tab:IMFEMV11H05} shows the IMF-weighted wind yields (top), calculated as, 
\begin{equation}\label{imfeq}
    m^{\rm IMF}_{i} = m^{\rm{wind}}_{im} \times M^{-1.9}_{i}
\end{equation} for stars with initial masses ranging from 50-500\Mdot.
We find that VMS make a substantial contribution to the net yields, even when weighted by an IMF. For instance, when compared to the wind yields of lower mass (12-40\Mdot) stars from \cite{hirschi05} we find that the contribution of VMS winds results in 10 times more ejected mass of $^{14}$N than the wind contribution of O stars (12-60\Mdot). In fact, the positive $^{12}$C and $^{16}$O yields of 60-100\Mdot\ compared to the negative yields of these 12-40\Mdot\ stars. Table \ref{tab:IMFEMV11H05} demonstrates the predominant effect that VMS have on their host environment compared to standard O stars in terms of their IMF-weighted yields (top) and IMF-weighted ejected masses (bottom).
 
We find that our VMS models still contribute 10 times more mass than a 50\Mdot\ star when weighted by an IMF, since they lose $\sim$90\% of their mass over their lifetime they can contribute significant amounts of mass back into their host galaxy, we refer back to the total ejected masses of Table \ref{tab:EMyieldsV11}. Furthermore, while the 50\Mdot\ stars may be $\sim$ 100 times more abundant, they eject $\sim$10 times less mass of $^{4}$He, $^{12}$C and $^{16}$O than the IMF-weighted VMS, suggesting that VMS ejecta are a significant provider of enriched material to their environments. Interestingly, the transition point model with M$_{\rm{init}}$ $=$ 80\Mdot\ ejects more $^{12}$C than the 100-500\Mdot\ stars. We also note that while the contribution of SNe ejecta of lower mass stars will play a role at t $>$ 6Myr, for young clusters, the ejected masses and net yields will be dominated by winds from VMS for t$\approx$ 0-4Myr.



\section{Conclusions}\label{conclusions}
The stellar wind contribution of VMS is investigated in this work, with enhanced mass-loss rates appropriate for stars which are observed to have optically thick winds. We have provided a comparison of the resulting ejected masses and net wind yields when implementing these enhanced winds with the previously adopted standard O star winds. We present the nucleosynthesis of VMS throughout their evolution, from core H-burning until core O-exhaustion, calculated with a large nuclear network comprising 92 isotopes. The dominant effects are explored during the MS evolution, with consequences for the post-MS. We consider the impact of stellar winds in lower Z environments for a subset of models. Finally, we evaluate the contribution of VMS winds compared with standard O stars, with IMF-weighted yields and a comparison of ejected masses for M$=$ 50\Mdot\ and M $>$ 100\Mdot. 

On the MS, 95\% of the total wind yields are produced, compared to just 5\% of the total wind yields which are ejected on the post-MS. This showcases the dominance of MS winds of VMS when compared to evolved stars. We compare the effects of enhanced, optically thick winds as opposed to standard, optically thin winds on the MS and find that VMS with enhanced winds eject up to 10 times more H-burning products of $^{14}$N, $^{20}$Ne, $^{23}$Na, and $^{26}$Al than VMS with standard winds. 

During the entire evolution of VMS, enhanced winds yield 10 times more $^{14}$N than M$\lesssim$ 50\Mdot\ O stars, but more importantly they yield positive amounts of $^{4}$He $^{12}$C, and $^{22}$Ne, relative to their initial abundances when compared to O stars which do not replenish their environments via stellar winds. In fact, single stars with initial masses below 50\Mdot\ do not eject any $^{26}$Al, but VMS eject 10$^{-2}$ to 10$^{-3}$ of $^{26}$Al and are likely responsible for the significant mass of $^{26}$Al observed in our Galaxy. 

Moreover, we show that a 100\Mdot\ star with enhanced winds ejects 100 times more $^{12}$C, $^{16}$O, and $^{28}$Si, than a 50\Mdot\ star. We also find that the ejected masses of $^{20}$Ne, $^{23}$Na and $^{26}$Al increase with increasing initial mass. This suggests that the presence of H-products such as $^{20}$Ne, $^{23}$Na and $^{27}$Al, seen in globular clusters \citep{gratton04, bastianlardo18} steers towards VMS. Although our models are computed for \Zdot, whilst globular clusters typically have low Z ([Fe/H] $=$ -1.5), there is no reason the nucleosynthetic production of Na or Al would be affected. Whether this material is ejected in sufficient quantities to explain the observed anti-correlations in globular clusters remains to be seen. While mass-loss rates are expected to be reduced at lower Z, and VMS would normally not be considered to be sufficiently numerous to pollute globular clusters with sufficient material, we have shown here that the total mass loss is an order of magnitude higher than previously considered. Therefore, for more definitive answers on the role of VMS as polluters in globular clusters we need to consider VMS models with appropriate mass-loss scalings, such as the recent low-Z mass-loss framework of Sabhahit et al. (2023). Interestingly, we discover that the intermediate mass range of transition stars with M$_{\rm{init}}$ $=$ 80-100\Mdot\ eject more $^{12}$C and $^{16}$O than higher mass stars as they do not experience CHE, or do not lose a significant amount of $^{4}$He before it is processed on the post-MS. Additionally, we show that VMS (M$\geq$ 100\Mdot) produce the same He-ZAMS mass and surface composition regardless of initial mass, when implementing enhanced stellar winds. 

Finally, when weighted by an IMF, uncovering the realistic contribution of VMS, we find that a 100\Mdot\ with enhanced winds still ejects 10 times more $^{4}$He, $^{12}$C, $^{14}$N, $^{16}$O  and $^{20,22}$Ne, than a 50\Mdot\ star.
Our conclusions reflect the significant impact that VMS winds have on their host galaxy or young cluster. The vast amount of mass lost already on the MS illustrates the presiding role that VMS winds have in replenishing their environments with reprocessed material, such as $^{4}$He, $^{12}$C, $^{14}$N, and $^{16}$O. In fact, adopting the appropriate wind prescription for VMS in both stellar evolution calculations and GCE simulations is crucial for providing accurate yields and ejected masses, as well as impacting many other stellar and galactic properties as highlighted throughout this work. We find that by adopting enhanced winds or standard O star winds, that the ejected masses and post-MS stellar masses can differ by $\gtrsim$ 100\Mdot. Finally, the winds of VMS prove to be the dominant source of $^{26}$Al, with 10-100 times more mass ejected by VMS than O stars or evolved WR stars, in a given population showcasing that VMS may be responsible for the enrichment of observed $^{26}$Al in the Galaxy and should be considered in future work.

\section*{Acknowledgements}
The authors acknowledge MESA authors and developers for their continued revisions and public accessibility of the code. JSV, AML, and ERH are supported by STFC funding under grant number ST/V000233/1 in the context of the BRIDGCE UK Network. RH acknowledges support from STFC, the World Premier International Research Centre Initiative (WPI Initiative), MEXT, Japan and the IReNA AccelNet Network of Networks (National Science Foundation, Grant No. OISE-1927130). This article is based upon work from the ChETEC COST Action (CA16117) and the European Union’s Horizon 2020 research and innovation programme (ChETEC-INFRA, Grant No. 101008324).
\section*{Data Availability}
The data underlying this article will be shared on reasonable request
to the corresponding author.
\typeout{}

\bibliographystyle{mnras}
\bibliography{Ref2.bib}
\begin{appendix}
\section{Main Sequence net yields}
\begin{table*}
    \centering
    \begin{tabular}{c|c|c|c|c|c|c|c|c|c|c|c|c|c|c|c}
    \hline
        $M_{\rm{i}}/\rm{M}_{\odot}$  & $^{4}$He & $^{12}$C & $^{14}$N & $^{16}$O & $^{20}$Ne & $^{22}$Ne & $^{23}$Na &$^{24}$Mg&$^{26}$Mg &$^{26}$Al &$^{27}$Al & $^{28}$Si \\
        \hline \hline
50&		5.86E-7&	-7.66E-7&	3.55E-8&	-1.23E-10&	-2.55E-11&	-2.91E-11	&2.78E-11	&-7.65E-12	&-1.15E-12&	0&	-5.99E-13&	-9.08E-12\\
80&	5.57&	-3.84E-2&	1.63E-1&	-1.35E-1&	-1.06E-3&	-2.80E-3	&4.13E-3	&2.66E-6&	3.65E-4&	7.09E-4&	1.88E-4&	1.16E-6\\
100&	2.06E+1&	-8.77E-2&	3.81E-1&	-3.18E-1&	-4.05E-3&	-6.34E-3&	1.14E-2&	1.69E-5&	-5.13E-5&	2.27E-3&	9.00E-4&	5.95E-6\\
200&	5.28E+1&	-2.60E-1&	1.13&	-9.45E-1&	-1.14E-2&	-1.87E-2	&3.30E-2	&5.04E-5	&-1.01E-3&	7.39E-3&	2.90E-3&	1.82E-5\\
300		&7.58E+1&	-4.35E-1&	1.88	&-1.57	&-1.72E-2&	-3.09E-2&	5.28E-2&	8.16E-5	&-2.22E-03	&1.27E-2&	4.73E-3	&2.85E-5\\
400&	9.59E+1&	-6.08E-1&	2.62	&-2.18	&-2.26E-2&	-4.29E-2&	7.17E-2	&1.13E-4&	-3.56E-3&	1.79E-2	&6.52E-3&	3.84E-5\\
500	&1.15E+2&	-7.81E-1&	3.36&	-2.80&	-2.77E-2&	-5.46E-2&	9.03E-2	&1.46E-4&	-4.97E-3	&2.30E-2&	8.33E-3	&4.80E-5\\
      \hline
    \end{tabular}
    \caption{Net yields calculated with equation \eqref{yieldeq} for V11 models during core H-burning only. Initial masses and yields are provided in solar mass units.}
    \label{tab:appV11MS}
\end{table*}

\section{Additional figures}
\begin{figure}
    \centering
    \includegraphics[width = \columnwidth]{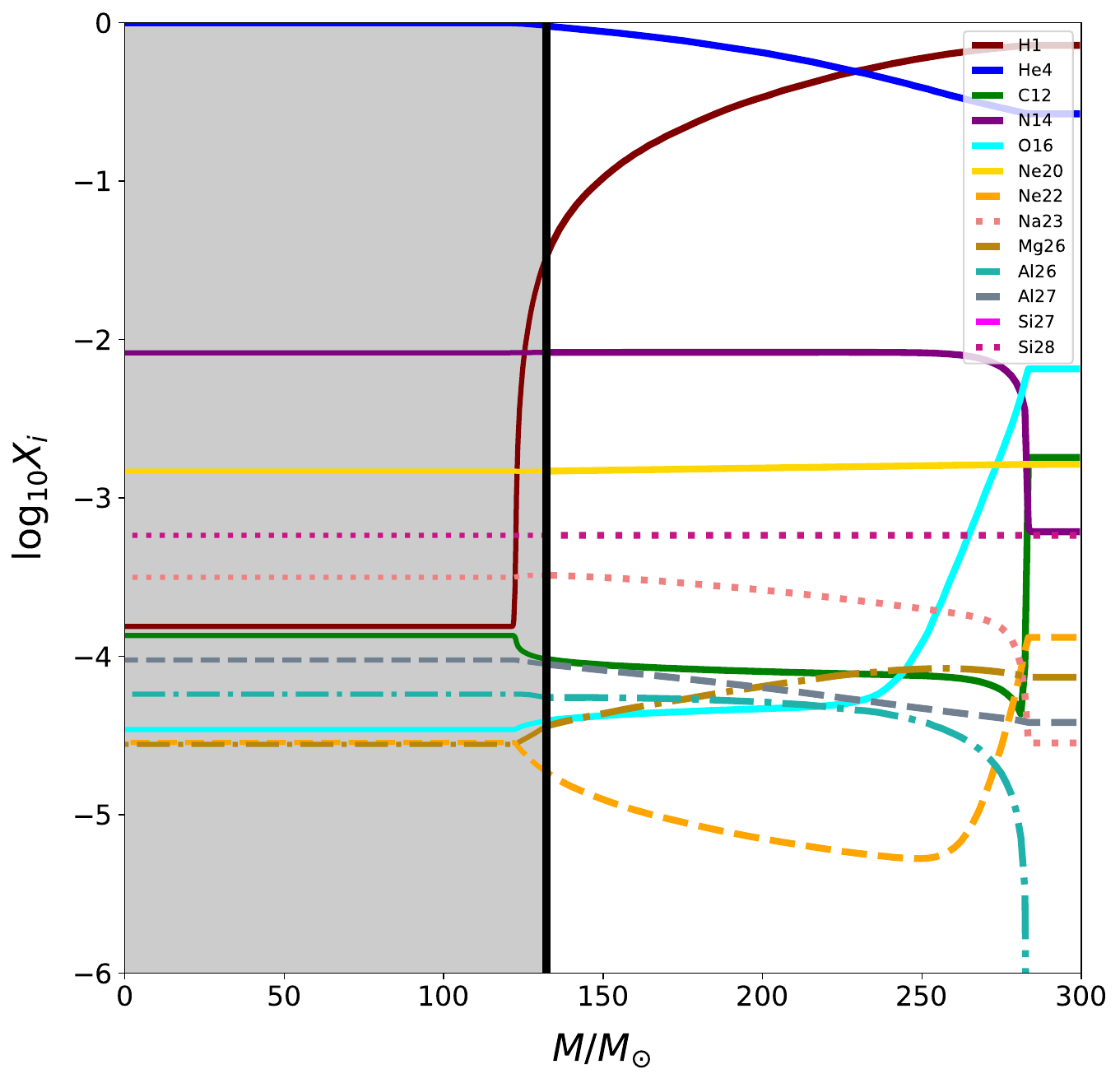}
    \caption{Time evolution of the surface composition during core H-burning, presented in log-scale represented as a function of stellar mass, with the interior composition shown in they grey shaded region, for a model with an initial mass of 300\Mdot\ implementing the V01 wind prescription, as in Fig.\ref{fig:300S22TAMS}.}
    \label{fig:300V01TAMS}
\end{figure}
\begin{figure}
    \centering
    \includegraphics[width = \columnwidth]{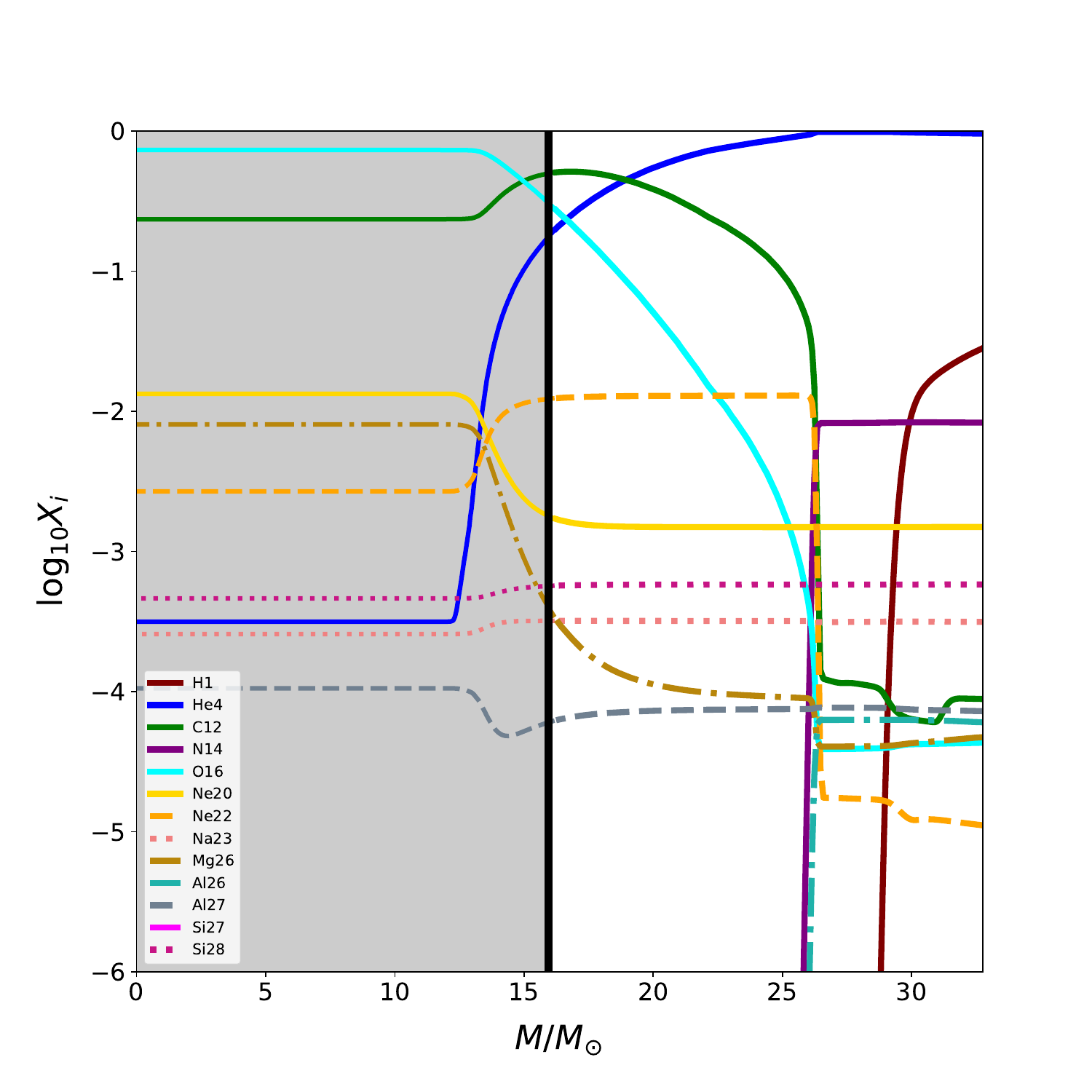}
    \caption{Time evolution of the surface composition during core He-burning for a 100\Mdot\ model at \Zdot. Isotopes are shown in log-scale as a function of stellar mass with the interior composition shown at the end of core He-burning (grey shaded region), as in Fig. \ref{fig:300HeTAMS}, applying the V11 wind prescription. }
    \label{fig:yield100HeTams}
\end{figure}

\begin{figure}
    \centering
    \includegraphics[width = \columnwidth]{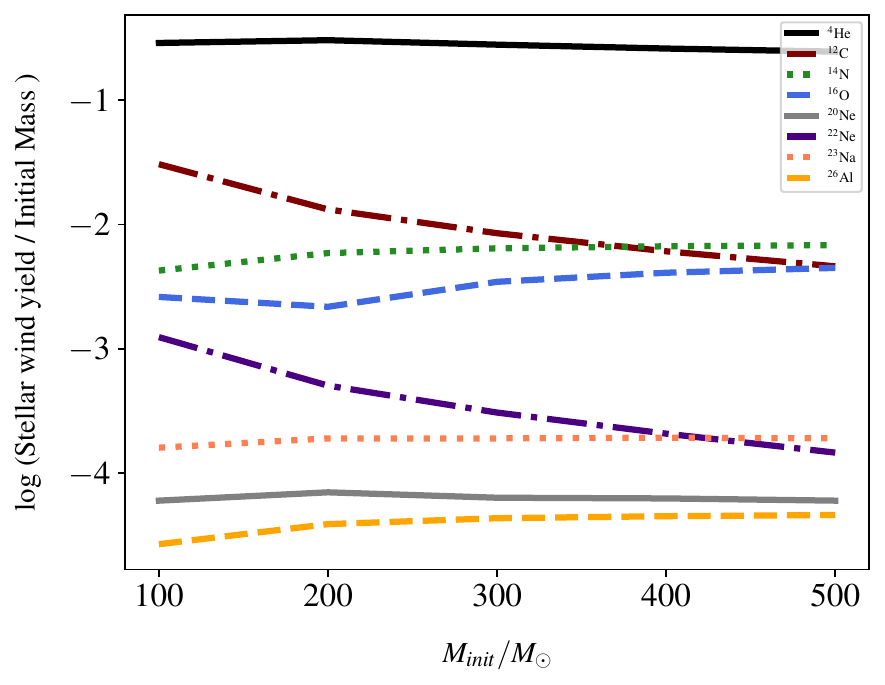}
    \caption{Stellar wind yields presented logarithmically as a fraction of initial mass and shown as a function of initial mass for models applying the V11 wind prescription. }
    \label{fig:Mini_yield}
\end{figure}
\begin{figure*}
    \centering
    \includegraphics[width = \textwidth]{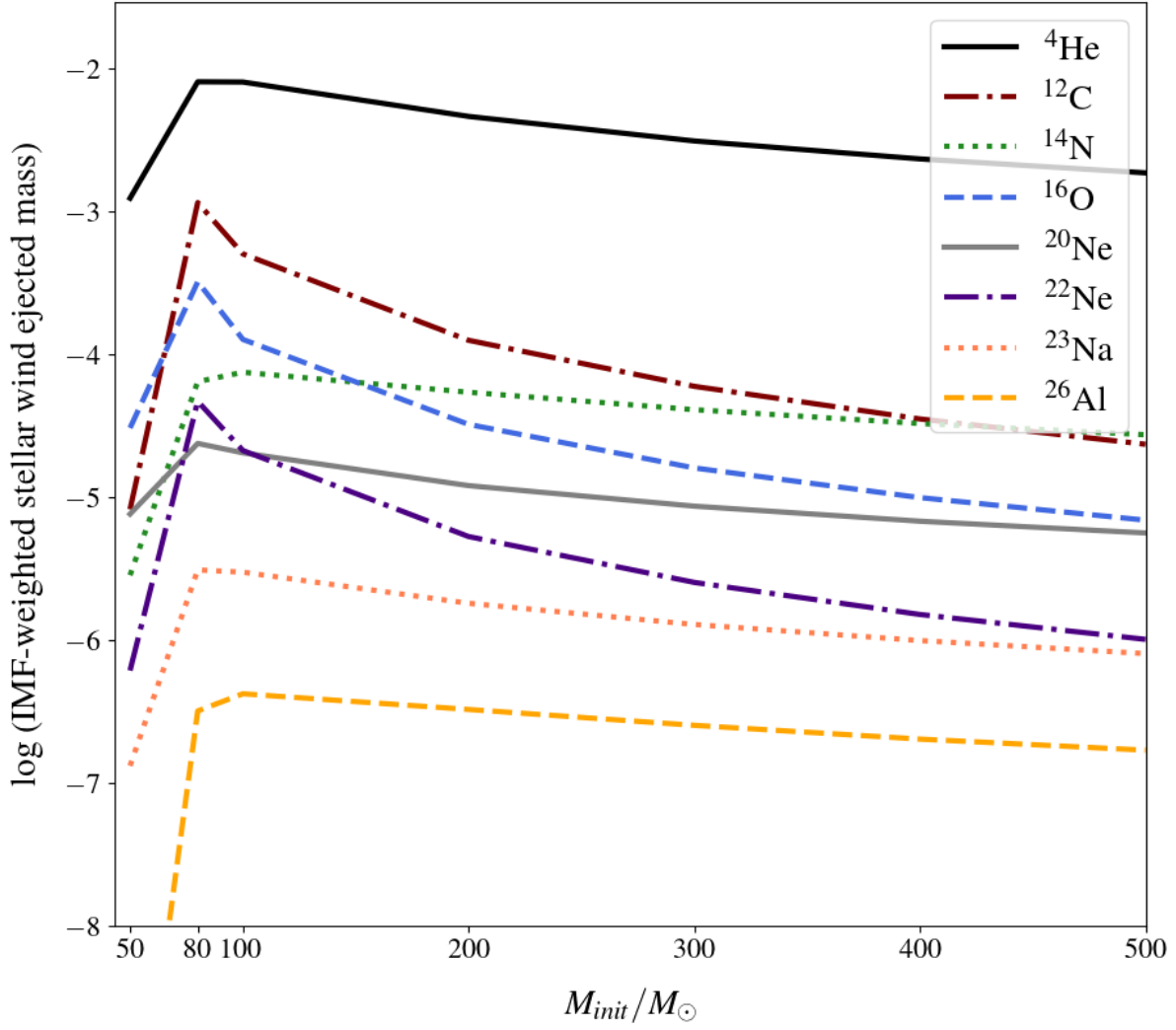}
    \caption{IMF-weighted wind ejected masses as a function of initial mass, as shown in Fig. \ref{fig:salpeterIMF} but with an IMF adopted from \citet{Schneider18} where $M^{-1.90}$.}
    \label{fig:IMFschneider}
\end{figure*}
\end{appendix}
\end{document}